\documentclass[onecolumn]{aa}

\usepackage{amsmath,amssymb}
\usepackage{graphicx}
\usepackage{txfonts}
\usepackage{natbib}

\begin{document}

\title{Theoretical formulation of Doppler redistribution in \\ 
scattering polarization within the framework of the \\
velocity-space density matrix formalism}

\titlerunning{Velocity density matrix correlations and Doppler redistribution}

\author{
L. Belluzzi\inst{\ref{inst1},\ref{inst2}}
\and E. Landi Degl'Innocenti\inst{\ref{inst3}}
\and J. Trujillo Bueno\inst{\ref{inst1},\ref{inst2},\ref{inst4}}
}

\authorrunning{Belluzzi, Landi Degl'Innocenti, \& Trujillo Bueno}

\institute{
Instituto de astrof\'isica de Canarias, C. V\'ia L\'actea s/n, E-38205 La 
Laguna, Tenerife, Spain\label{inst1}
\and
Departamento de Astrof\'isica, Facultad de F\'isica, Universidad de La Laguna, 
E-38200 La Laguna, Tenerife, Spain\label{inst2}
\and
Dipartimento di Fisica e Astrofisica, Universit\`a di Firenze, 
Largo E. Fermi 2, I-50125 Firenze, Italy\label{inst3}
\and
Consejo Superior de Investigaciones Cient\'ificas, Spain\label{inst4}
}

\abstract{
Within the framework of the density matrix theory for the generation and 
transfer of polarized radiation, velocity density matrix correlations represent 
an important physical aspect that, however, is often neglected in practical 
applications by adopting the simplifying approximation of complete 
redistribution on velocity.
In this paper, we present an application of the Non-LTE problem for polarized 
radiation taking such correlations into account through the velocity-space 
density matrix formalism.
We consider a two-level atom with infinitely sharp upper and lower levels, and 
we derive the corresponding statistical equilibrium equations neglecting the 
contribution of velocity-changing collisions. 
Coupling such equations with the radiative transfer equations for polarized 
radiation, we derive a set of coupled equations for the velocity-dependent 
source function.
This set of equations is then particularized to the case of a plane-parallel 
atmosphere.
The equations presented in this paper provide a complete and solid description 
of the physics of pure Doppler redistribution, a phenomenon generally described 
within the framework of the redistribution matrix formalism.
The redistribution matrix corresponding to this problem (generally referred to 
as $R_I$) is derived starting from the statistical equilibrium equations for 
the velocity-space density matrix, and from the radiative transfer equations 
for polarized radiation, thus showing the equivalence of the two approaches.
}


\maketitle

\section{Introduction}
When polarization phenomena are considered, the usual description of the 
excitation state of an atomic system in terms of the population of its energy 
levels is not adequate, and it is necessary to specify the population of 
each magnetic sublevel, as well as the quantum interference (or coherence) that 
might be present between pairs of them.
Whenever the magnetic sublevels of a given energy level are unevenly populated
and/or quantum interference between pairs of them are present, the atomic 
system is said to be polarized.
Atomic polarization is generally induced whenever an atomic system is excited 
by means of a physical process which is not spatially isotropic.

A powerful theoretical tool that allows to describe in a very compact way 
the full excitation state of an atomic system is the so-called density 
operator \citep[see][]{Fan57}.
The most natural basis for defining the matrix elements of the density operator 
is the basis of the eigenvectors of the total angular momentum 
$| \alpha J M \rangle$, with $J$ the total angular momentum, $M$ its 
projection along the quantization axis, and $\alpha$ a set of inner quantum 
numbers. On this basis, the elements of the density matrix are given by
\begin{equation}
	\langle \alpha J M | \, \hat{\rho} \, |\alpha^{\prime} J^{\prime} 
	M^{\prime} \rangle \equiv \rho(\alpha J M, \alpha^{\prime} J^{\prime} 
	M^{\prime}) \; ,
\end{equation}
with $\hat{\rho}$ the density operator.
The diagonal elements represent the populations of the magnetic sublevels,
the off-diagonal elements the quantum interference (or coherence) between 
different magnetic sublevels \citep[see][hereafter LL04]{Lan04}.
For the sake of simplicity, in this paper we will only consider interference 
between pairs of magnetic sublevels pertaining to the same $J$-level, which is 
a good approximation for the investigation of many solar spectral lines 
\citep[see LL04 and][for a detailed discussion on the importance of 
interference between different $J$-levels in determining the wing polarization 
of multiplet lines]{Bel11a}.
We will thus consider only the density matrix elements of the form 
$\rho(\alpha J M, \alpha J M^{\prime})$.
In general it is convenient to work in terms of the multipole moments of the 
density matrix (or spherical statistical tensors)
\begin{equation}
	\rho^K_Q(\alpha J) = \sum_{M M^{\prime}} (-1)^{J-M} \sqrt{2K + 1}
	\, \left( \begin{array}{ccc}
		J & J & K \\
		M & -M^{\prime} & -Q 
	\end{array} \right) \,
	\rho(\alpha J M, \alpha J M^{\prime}) \; ,
\end{equation}
which transform as irreducible tensors under a rotation of the reference 
system.

Since the radiation field experienced by an atom depends, because of the 
Doppler effect, on its velocity, the density matrix will in general depend 
on the velocity $\boldsymbol{\varv}$ of the atom.
Indicating with $\rho^K_Q(\alpha J; \boldsymbol{\varv})$ the 
$\boldsymbol{\varv}$-dependent spherical statistical tensor of the atomic 
system, and with $f(\boldsymbol{\varv})$ the velocity distribution function of 
the atoms in a given point of the plasma, a complete statistical description of 
the atom is given by the product $f(\boldsymbol{\varv}) \, \rho^K_Q(\alpha J; 
\boldsymbol{\varv})$, generally referred to as {\it velocity-space density 
matrix}.
The need for the introduction of this quantity was put forward by \citet{Lan96}
who pointed out the importance of velocity density matrix correlations in 
polarized radiative transfer.
The statistical equilibrium equations for the velocity-space density matrix 
can be written in the form (see Section 13.2 of LL04)
\begin{equation}
   \frac{\rm d}{{\rm d} t} \left[ f(\boldsymbol{\varv}) \,
   \rho^K_Q(\alpha J; \boldsymbol{\varv}) \right] = f(\boldsymbol{\varv}) 
   \left( \frac{\rm d}{{\rm d} t} \, \rho^K_Q(\alpha J;\boldsymbol{\varv}) 
   \right)_{0} + \left( \frac{\delta}{{\delta} t} 
   \left[ f(\boldsymbol{\varv}) \, \rho^K_Q(\alpha J; \boldsymbol{\varv}) 
   \right] \right)_{\rm vel. \, chang. \, coll.} \; .
\label{Eq:Gen_SEE}
\end{equation}
The first term in the right-hand side is due to processes which, as a first 
approximation, are not effective in changing the velocity of the atom: these 
include radiative processes (absorption and emission of photons), inelastic 
and superelastic collisions with electrons, and depolarizing collisions with 
neutral hydrogen atoms.\footnote{Depolarizing collisions are elastic collisions 
due to long-range interactions, and are thus ineffective in changing 
appreciably the velocity of the atom.}
The second term, which can be regarded as a generalization of the Boltzmann 
term which is met in the kinetic theory of gases, is due to collisions that
are able to modify the velocity of the atom (velocity-changing collisions). 
Such collisions, which are characterized by very small impact parameters and by 
rather large exchange of kinetic energy, generally induce transitions between 
different energy levels, and thus affect the atomic density matrix.
Because of the generalized Boltzmann term, Equation~(\ref{Eq:Gen_SEE}) is 
extremely complicated, and two different approximations are generally 
introduced.

The first one consists in neglecting the generalized Boltzmann term 
({\it velocity-coherence approximation}).  
As discussed in Section 13.2 of LL04, this is a good approximation in the outer 
layers of a stellar atmosphere, where the number density of perturbers 
responsible for velocity-changing collisions (typically hydrogen atoms or ions) 
is sufficiently low, and this kind of collisions are indeed negligible.
Under this approximation, the $\boldsymbol{\varv}$-dependence of the density 
matrix is only due to the fact that atoms moving with different velocities may 
experience, because of the Doppler effect, different radiation fields.

The second approximation is to assume that velocity-changing collisions 
are so efficient in reshuffling the atomic velocities that any velocity 
density matrix correlation is lost ({\it complete redistribution on velocity 
approximation}).
In this case, the velocity-space density matrix is given by 
$f(\boldsymbol{\varv}) \, \rho^K_Q(\alpha J)$, the density matrix being 
independent of $\boldsymbol{\varv}$. 
Since velocity-changing collisions also contribute to depolarize the atomic 
system, when the complete redistribution on velocity approximation is 
justified, polarization phenomena are generally negligible.
For this reason, it is customary to consider an intermediate approach which 
consists in neglecting the generalized Boltzmann term, still assuming a 
velocity-independent density matrix.

The general problem of interpreting the spectropolarimetric profiles of lines 
formed in an optically thick plasma, such as a stellar atmosphere, requires the 
self-consistent solution of the statistical equilibrium equations and of the 
radiative transfer equations, taking into account polarization phenomena, both 
in the atomic system and in the radiation field \citep[see][]{JTB03}.
This problem has been referred to as the Non-LTE problem of the $2^{\rm nd}$ 
kind (see LL04), so to distinguish it from the usual Non-LTE problem where 
polarization phenomena are neglected. 
A detailed discussion of the general Non-LTE problem of the $2^{\rm nd}$ kind, 
under the approximation of complete redistribution on velocity, is presented in 
Chapter~14 of LL04.

In this paper, we present an application of the Non-LTE problem of the 
$2^{\rm nd}$ kind under the velocity-coherence approximation previously 
discussed.
We consider the basic case of a two-level atom with infinitely sharp upper and 
lower levels.
Starting from the statistical equilibrium equations for the velocity-space 
density matrix (Sect.~3), and from the radiative transfer equations for 
polarized radiation (Sect.~4), we derive a set of coupled equations for the 
velocity-dependent source function (Sect.~5). 
This set of coupled equations is then specified to the particular case of a 
Maxwellian distribution of velocities, and it is finally applied to the 
particular case of a plane-parallel atmosphere (Sect.~6).
From the same equations we also derive the redistribution phase-matrix 
corresponding to this physical problem (generally referred to as $R_{I}$), 
thus showing the equivalence of this latter, widely applied approach to the 
one described in this paper (Sect.~7).

The more realistic case of partial frequency redistribution, assuming a 
two-level model atom with infinitely-sharp and unpolarized lower level, and 
with a naturally and/or collisionally broadened upper level, has been 
considered by several authors \citep[e.g.,][]{Omo72,Omo73,Dom88,Bom97a,Bom97b,
Sam12}, providing expressions for the $R_{II}$ and $R_{III}$ redistribution 
matrices.
In this paper we show, for the academic case of a two-level atom with 
infinitely-sharp upper and lower levels, how the phenomenon of pure Doppler 
redistribution can be rigorously described through the velocity-space density 
matrix formalism.
As we will see, this formalism allows to describe the physics of the 
atom-photon interaction in a very transparent way (e.g., it allows to clearly 
identify correlations between atoms located at different points of the plasma), 
and it highlights the underlying approximations (e.g., the neglect of 
velocity-changing collisions).
The equations that are obtained are very general, and can be applied to 
arbitrary velocity distributions. Moreover, they are able to account for the 
presence of atomic polarization in the lower level of the considered 
transition, and they are perfectly suitable for a generalization to the 
multilevel case.
The application of the velocity-space density matrix formalism for the 
description of a two-level atom with broadened upper level is presently under 
investigation, and will not be discussed in this paper.

\section{Formulation of the problem: hypotheses and approximations}
We consider a two-level atom without hyperfine structure and with 
unpolarized lower level. 
Consistently with the last assumption, we suppose that the radiation field 
incident on the atom is weak, in the sense that the average number of photons 
per mode, $\bar n$, is much smaller than unity, which justifies to neglect 
stimulated emission.

We suppose that a collection of such atoms is distributed within a static
medium of arbitrary shape. 
In this medium the atoms interact with a magnetic field, $\vec B$, and with a 
population of colliding particles having a Maxwellian distribution of 
velocities characterized by the temperature $T$.
No restriction is made on the spatial variation within the medium of the 
temperature $T$ of the colliders, of the densities of the atoms and colliders,
and of the magnetic field vector $\vec B$.
We suppose that the magnetic field is weak (in the sense that the associated 
Larmor frequency $\nu_{\rm L}$ is much smaller than the frequency width
$\Delta \nu_{\rm P}$ of the absorption profile) and that the inverse lifetime 
of the upper level, $\gamma_u$, is also much smaller than $\Delta \nu_{\rm P}$, 
so that the flat-spectrum approximation is satisfied.\footnote{The 
applicability of the flat-spectrum approximation follows from the two 
inequalities $\gamma_u \ll \Delta \nu_{\rm P}$ and $\nu_{\rm L} \ll \Delta 
\nu_{\rm P}$. 
The latter obviously implies an upper limit on the magnetic field intensities 
that can be handled by this formalism.}

The atoms are characterized by an arbitrary velocity distribution 
$f(\boldsymbol{\varv})$ that, for the sake of simplicity, we assume to be 
constant throughout the medium.
Taking velocity density matrix correlations into account, at any point P of 
the medium, of coordinate $\vec{x}$, the atom is thus described by the 
velocity-space density matrix $f(\boldsymbol{\varv}) \left[ \rho^K_Q(\alpha J; 
\boldsymbol{\varv}) \right]_{\vec{x}}$, where $(\alpha J) \! = \! 
(\alpha_u J_u)$ for the upper level and $(\alpha J)\! =\! (\alpha_\ell J_\ell)$ 
for the lower level.
Neglecting velocity-changing collisions (i.e. assuming the velocity-coherence 
approximation described in Sect.~1), the velocity-space density matrix evolves 
with time according to the equation
\begin{equation}
	\frac{\rm d}{{\rm d} t} \left( f(\boldsymbol{\varv}) 
	\left[ \rho^K_Q(\alpha J; \boldsymbol{\varv}) \right]_{\vec x} 
	\right) = f(\boldsymbol{\varv}) \left( \frac{\rm d}{{\rm d} t} 
	\left[ \rho^K_Q(\alpha J; \boldsymbol{\varv}) \right]_{\vec x} 
	\right)_{0} \; ,
\end{equation}
which is solved by
\begin{align}
	\frac{\rm d}{{\rm d} t} \, f(\boldsymbol{\varv}) = & \, 0  \; , \\
	\frac{\rm d}{{\rm d} t} \, 
	\left[ \rho^K_Q(\alpha J; \boldsymbol{\varv}) \right]_{\vec x} = & \,
	\left( \frac{\rm d}{{\rm d} t} 
	\left[ \rho^K_Q(\alpha J; \boldsymbol{\varv}) \right]_{\vec x} 
	\right)_{0} 
	\; .
	\label{Eq:SEE_rho}
\end{align}
As previously pointed out, the term in the right-hand-side of 
Eq.~(\ref{Eq:SEE_rho}) contains the ``ordinary'' processes due to the 
atom-radiation interaction and to collisions. 
The value of $\left[ \rho^K_Q(\alpha J; \boldsymbol{\varv}) \right]_{\vec x}$ 
can thus be found by solving the statistical equilibrium equations presented 
in LL04, taking properly into account the explicit dependence of the radiative 
and collisional rates on the velocity of the atom.
The velocity distribution $f(\boldsymbol{\varv})$, on the other hand, remains 
undetermined, and can only be established by means of different physical 
considerations.
In many cases, it can simply be assumed to be a Maxwellian, possibly centered 
at a non-zero velocity, like in the case of the solar wind.
In the next Sections, if not explicitly specified, we consider an arbitrary 
velocity distribution $f(\boldsymbol{\varv})$.

Finally, we neglect the broadening effect of elastic collisions, so that 
both the upper and lower level of the atom can be considered infinitely sharp.
Consequently, we also neglect any frequency redistribution effect due to 
elastic collisions.
In terms of scattering processes, our model thus allows us to describe coherent 
scattering in the atom rest frame, with purely Doppler redistribution in the 
observer frame.
Within the framework of the redistribution matrix formalism, following the 
terminology introduced by \citet{Hum62}, this kind of process, in the case of a 
Maxwellian velocity distribution, is described by the $R_I$ redistribution 
function.

\section{The Statistical Equilibrium Equations}
Referring to the geometry of figure 1, in the `fixed' (or laboratory) reference 
system $\Sigma$, the statistical equilibrium equation for the multipole moments 
of the upper level is the same as Eq.~(14.2) of LL04, with the only difference 
that the radiative and collisional rates now depend explicitly on the velocity 
$\boldsymbol{\varv}$ of the atom.
We thus have
\begin{equation}
\begin{split}
   \left( \frac{\rm d}{{\rm d} t} \left[ \rho^K_Q(\alpha_u J_u ; 
   \boldsymbol{\varv} ) \right]_{\vec x} \right)_{0} = &
   -2 \pi \, {\rm i} \,\nu_{\rm L} \, g_{\alpha_u J_u} \, \sum_{Q'} \, 
   {\mathcal K}^K_{QQ'} \, \left[ \rho^K_{Q'}(\alpha_u J_u ; 
   \boldsymbol{\varv} ) \right]_{\vec x} \\
   & + \sum_{K'Q'} \, {\mathbb T}_{\! \rm A}(\alpha_u J_u KQ, \alpha_\ell 
   J_\ell K'Q') \, \left[ \rho^{K'}_{Q'}(\alpha_\ell J_\ell ; 
   \boldsymbol{\varv} ) \right]_{\vec x} \\
   & - \sum_{K'Q'} \, \left[ {\mathbb R}_{\rm E}(\alpha_u J_u KQK'Q') 
   + {\mathbb R}_{\rm S}^{\phantom K}(\alpha_u J_u KQK'Q') \right]
   \left[ \rho^{K'}_{Q'}(\alpha_u J_u ; \boldsymbol{\varv}) \right]_{\vec x} \\
   & + \, \sqrt{\frac{2J_\ell+1}{2J_u+1}} \, 
   C_{\rm I}^{(K)}(\alpha_u J_u, \alpha_\ell J_\ell) \,
   \left[ \rho^K_Q(\alpha_\ell J_\ell ; \boldsymbol{\varv}) \right]_{\vec x} \\
   & - \left[ C_{\rm S}^{(0)}(\alpha_\ell J_\ell, \alpha_u J_u) \,
   + D^{\, (K)}(\alpha_u J_u) \right]
   \left[ \rho^K_Q(\alpha_u J_u ; \boldsymbol{\varv} ) \right]_{\vec x} \;\; ,
\label{Eq:SEE}
\end{split}
\end{equation}
where all the rates are evaluated at point $\vec x$, $\nu_{\rm L}$ is the
Larmor frequency at the same point, and $g_{\alpha_u J_u}$ is the Land\'e 
factor of the upper level.
\begin{figure}[!t]
\centering
\includegraphics[width=0.5\textwidth]{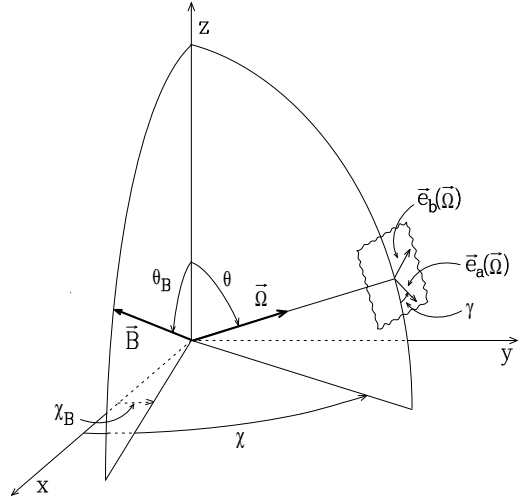}
\caption{At each point {\rm P} of the medium, the magnetic field vector 
$\vec{B}$ is specified by the angles $\theta_B$ and $\chi_B$, defined in the
fixed reference system $\Sigma \equiv (xyz)$.
The polarization unit vector $\vec e_a(\vec\Omega)$ specifies the reference 
direction for positive $Q$ of the radiation flowing through P in the direction 
$\vec\Omega$, specified by the angles $\theta$ and $\chi$ in the fixed 
reference system.}
\end{figure}
The kernel ${\mathcal K}^K_{QQ'}$ is given by (see Eq.~(7.79) of LL04)
\begin{equation}
	{\mathcal K}^K_{QQ^{\prime}} = \sum_{Q^{\prime \prime}} \, 
	{\mathcal D}^K_{Q^{\prime \prime} Q}(R_B)^{\ast} \; 
	Q^{\prime \prime} \; {\mathcal D}^K_{Q^{\prime \prime} Q^{\prime}}
	(R_B) \;\; ,
\end{equation}
where ${\mathcal D}^K_{QQ^{\prime}}(R)$ are rotation matrices, and $R_B$ is 
the rotation that carries the local `magnetic' reference system (having the 
$z$-axis aligned with the magnetic field) into the `fixed' reference system 
$\Sigma$. 
In terms of Euler angles one simply has (see Fig.~1 for the definition of the 
angles)
\begin{equation}
   R_B \equiv (-\gamma_B,-\theta_B,-\chi_B) \nonumber \;\; ,
\end{equation}
where $\gamma_B$ is an arbitrary angle that can be set to zero. 
The main properties and the explicit expressions of the components of 
${\mathcal K}^K_{QQ^{\prime}}$ are given in App.~19 of LL04.
The assumptions that we have introduced yield two basic simplifications in 
Eq.~(\ref{Eq:SEE}):
\begin{itemize}
	\item{because stimulation effects are neglected, the relaxation rate
		${\mathbb R}_{\rm S}$ is zero;}
	\item{because lower-level polarization is neglected, the statistical
		tensors of the lower level reduce to 
		$\rho^K_Q(\alpha_\ell J_\ell ; \boldsymbol{\varv} ) = 
		\rho^0_0(\alpha_\ell J_\ell ; \boldsymbol{\varv} ) \,
		\delta_{K0} \, \delta_{Q0}$.} 
\end{itemize}
This implies that the only radiative rate needed in Eq.~(\ref{Eq:SEE}) is the 
absorption rate, ${\mathbb T}_{\! \rm A}(\alpha_u J_u KQ, \alpha_\ell J_\ell \, 
0 \, 0)$. 
Taking into account that the upper and lower levels of the atom are infinitely 
sharp, such rate, in the fixed reference frame, is given by (cf. Eq.~(10.9) of 
LL04)
\begin{equation}
   {\mathbb T}_{\rm A}(\alpha_u J_u K Q, \alpha_\ell J_\ell \, 0 \, 0) = 
   \sqrt{3 (2 J_\ell + 1)} \; B(\alpha_\ell J_\ell \rightarrow \alpha_u J_u) 
   \, (-1)^{1 + J_\ell + J_u + Q}
   \left\{ \!
   \begin{array}{ccc}
	   1 & 1 & K \\ 
	   J_u & J_u & J_\ell
   \end{array}
   \right\} \,
   \left[ J^K_{-Q}(\nu_0) \right]_{\rm c.f.} \;\, ,
\end{equation}
where $\left[ J^K_{-Q}(\nu_0) \right]_{\rm c.f.}$ is the radiation field tensor
calculated in the atom rest frame, or {\it comoving frame} (it describes the 
radiation field as ``seen'' by the atom, and it actually depends on its 
velocity $\boldsymbol{\varv}$ in the fixed reference frame), and where $\nu_0$ 
is the frequency of the transition between the upper and lower level of the 
atom.
In the comoving frame, the radiation field tensor is given by (see Eq.~(5.157) 
of LL04) 
\begin{equation}
   \left[ J^K_{-Q}(\nu_0) \right]_{\rm c.f.} = 
   \oint \frac{{\rm d} \Omega}{4 \pi} \, \sum_{i=0}^3 \,
   {\mathcal T}^K_{-Q}(i, \vec\Omega) \, 
   \left[ S_{\! i}(\nu_0, \vec\Omega) \right]_{\rm c.f.} \; ,
\end{equation}
where $\left[ S_{\! i}(\nu_0, \vec\Omega) \right]_{\rm c.f.}$ are the Stokes 
parameters of the radiation propagating along the direction $\vec\Omega$ at 
the frequency $\nu_0$, as defined in the comoving frame.
In the limit $\varv/c \ll 1$ aberration and further relativistic effects can 
be neglected and the radiation field tensor in the comoving frame can be 
calculated from the Stokes parameters in the fixed reference frame just taking 
into account the Doppler effect evaluated to first order in $\varv/c$. 
This brings to the following expression 
\begin{equation}
   \left[ J^K_{-Q}(\nu_0) \right]_{\rm c.f.} 
   = \oint \frac{{\rm d} \Omega}{4 \pi} \, \sum_{i=0}^3 \,
   {\mathcal T}^K_{-Q}(i, \vec\Omega) \; 
   S_{\! i} \left( \nu_0 + \nu_0 \frac{\boldsymbol{\varv} \cdot \vec\Omega}{c}, 
   \vec\Omega \right) \; ,
\label{Eq:JKQ_cf}
\end{equation}
where now $S_{\! i}$ are the Stokes parameters measured in the fixed frame.

As far as the the inelastic and superelastic collisional rates, 
$C_{\rm I}^{(K)} (\alpha_u J_u, \alpha_\ell J_\ell)$, and 
$C_{\rm S}^{(0)} (\alpha_\ell J_\ell,\alpha_u J_u)$ are concerned, we suppose 
that they do not depend on the velocity $\boldsymbol{\varv}$ of the atom 
undergoing the collision. 
This is a very good approximation, since such collisions are due to electrons, 
which travel with velocities much larger (typically by two orders of magnitude) 
than the velocity of the atom. 
Concerning depolarizing collisions (elastic collisions), basically due to 
neutral perturbers such as hydrogen atoms, the approximation can be more 
questionable. For this reason we will, from now on, denote such rates with 
the symbol $D^{(K)}(\alpha_u J_u ; \boldsymbol{\varv} )$.  

Taking into account these remarks, and recalling the expression of the 
radiative rate ${\mathbb R}_{\rm E}$ (cf. Eq.~(7.14e) of LL04), which 
obviously does not depend on the velocity of the atom, being due to spontaneous 
de-excitation processes, we can rewrite Eq.~(\ref{Eq:SEE}) in the form 
\begin{equation}
\begin{split}
   \left( \frac{\rm d}{{\rm d} t} \, \left[ 
   \rho^K_Q(\alpha_u J_u ; \boldsymbol{\varv} ) \right]_{\vec x} \right)_0 = &
   -2\pi \, {\rm i} \, \nu_{\rm L} \, g_{\alpha_u J_u} \, \sum_{Q'} \, 
   {\mathcal K}^K_{QQ'} \, \left[ \rho^K_{Q'}(\alpha_u J_u ; 
   \boldsymbol{\varv} ) \right]_{\vec x} \\ 
   & - \left[ A(\alpha_u J_u \rightarrow \alpha_\ell J_\ell)
   +C_{\rm S}^{(0)}(\alpha_\ell J_\ell, \alpha_u J_u) 
   +D^{(K)}(\alpha_u J_u ; \boldsymbol{\varv} ) \right] \left[ 
   \rho^K_Q(\alpha_u J_u ; \boldsymbol{\varv} ) \right]_{\vec x} \\
   & + \, \sqrt{\frac{2J_\ell + 1}{2J_u + 1}} \;
   \left[ B(\alpha_\ell J_\ell \rightarrow \alpha_u J_u) \,
   w^{(K)}_{J_u J_\ell} \; (-1)^Q \, \left[ J^K_{-Q}(\nu_0) \right]_{\rm c.f.} 
   + \delta_{K0} \, \delta_{Q0} \; 
   C_{\rm I}^{(0)}(\alpha_u J_u, \alpha_\ell J_\ell) \right] 
   \left[ \rho^0_0(\alpha_\ell J_\ell , \boldsymbol{\varv} ) 
   \right]_{\vec x} \; ,
\end{split}
\end{equation}
where the symbol $w^{(K)}_{J_u J_\ell}$ is given by Eq.~(10.11) of LL04.

Since the colliding particles have a Maxwellian distribution of velocities, 
we can apply the Einstein-Milne relation to connect the collisional rates due 
to inelastic and superelastic collisions (cf. Eq.~(10.49) of LL04). 
Next we divide both members by $A(\alpha_u J_u \rightarrow \alpha_\ell J_\ell)$ 
and introduce the usual notations (cf. Eqs.~(10.51) and (10.28) of LL04)
\begin{equation}
   \epsilon = \frac{C_{\rm S}^{(0)}(\alpha_\ell J_\ell, \alpha_u J_u)}
   {A(\alpha_u J_u \rightarrow \alpha_\ell J_\ell)} \, , \quad
   \delta^{(K)}_{\,u}(\boldsymbol{\varv} \,) = 
   \frac{D^{(K)}(\alpha_u J_u;\boldsymbol{\varv} \,)}
   {A(\alpha_u J_u \rightarrow \alpha_\ell J_\ell)} \, , \quad
   H_u = \frac{2\pi\nu_{\rm L} \; g_{\alpha_u J_u}}
   {A(\alpha_u J_u \rightarrow \alpha_\ell J_\ell)} \, .
\end{equation}
Recalling the relations between the Einstein coefficients (Eqs.~(7.8) of LL04),
we obtain, for stationary situations
\begin{equation}
\begin{split}
	\left[ 1 + \epsilon + \delta_{\,u}^{\, (K)}(\boldsymbol{\varv} \,) 
	\right] &
	\left[ \rho^K_Q(\alpha_u J_u ; \boldsymbol{\varv} ) \right]_{\vec x} 
	\, + \, {\rm i} \, H_u \sum_{Q'} \, {\mathcal K}^K_{QQ'} \,
	\left[ \rho^K_{Q'}(\alpha_u J_u ; \boldsymbol{\varv} ) 
	\right]_{\vec x} = \\ 
   	& = \frac{c^2}{2h\nu_0^3} \, \sqrt{\frac{2J_u+1}{2J_\ell+1}} \; 
   	\left[ w^{\, (K)}_{J_u J_\ell} \; (-1)^Q \, 
	[J^K_{-Q}(\nu_0)]_{\rm c.f.} 
   	+ \delta_{K0} \, \delta_{Q0} \; \epsilon \, B_T(\nu_0) \right] 
	\left[ \rho^0_0(\alpha_\ell J_\ell ; \boldsymbol{\varv} ) 
	\right]_{\vec x} \; ,
\end{split}
\end{equation}
where
\begin{equation}
   B_T(\nu_0)=\frac{2 h \nu_0^3}{c^2} \,
   {\rm exp} \left( -\frac{h \nu_0}{k_{\rm B}T} \right)
\end{equation}
is the Planck function in the Wien limit (consistently with the fact that 
stimulated emission is neglected).

In view of the following applications, it is convenient to rewrite the previous 
equation for the time evolution of the density matrix in a more compact form, 
by introducing suitable `source functions' for the different statistical 
tensors. 
Defining
\begin{equation}
	{\mathcal S}^K_Q(\boldsymbol{\varv}, \vec x \,)=\frac{2h \nu_0^3}{c^2}
   	\, \sqrt{\frac{2J_\ell+1}{2J_u+1}} \;\,
	\frac{\left[ \rho^K_Q(\alpha_u J_u; \boldsymbol{\varv} ) 
	\right]_{\vec x}}{\left[ \rho^0_0(\alpha_\ell J_\ell; 
	\boldsymbol{\varv} ) \right]_{\vec x}} \; ,    
\end{equation}
such equation becomes
\begin{equation}
	\left[ 1 + \epsilon + \delta_{\,u}^{\,(K)} (\boldsymbol{\varv} \,) 
	\right] {\mathcal S}^K_Q(\boldsymbol{\varv}, \vec x\,) + {\rm i} \, 
	H_u \sum_{Q'} \, {\mathcal K}^K_{QQ'} \, 
	{\mathcal S}^K_{Q'}(\boldsymbol{\varv}, \vec x\,)
   	= w^{\,(K)}_{J_u J_\ell} \; (-1)^Q \, \left[ J^K_{-Q}(\nu_0) 
   	\right]_{\rm c.f.} + \delta_{K0} \, \delta_{Q0} \; \epsilon \, 
   	B_T(\nu_0) \; .
\label{Eq:SF_SEE}
\end{equation}
The quantities ${\mathcal S}^K_Q(\boldsymbol{\varv}, \vec x \,)$ are the 
obvious generalization of the {\it irreducible components of the two-level atom
source function}, introduced in LL04. 
Now we also have an explicit dependence on $\boldsymbol{\varv}$.
This is because atoms having different velocities may have, in general, 
different source functions since, even at the same point in the medium, they 
experience, due to the Doppler effect, different radiation fields.

\section{The radiative transfer equation}
We consider now the radiative transfer equation. From Eq.~(6.83) of LL04
we have, neglecting stimulated emission
\begin{equation}
   \frac{\rm d}{{\rm d} s} \, S_{\! i}(\nu,\vec\Omega) = - \sum_{j=0}^3 \, 
   K_{ij}^{\, \rm A} \; S_{\!\! j}(\nu,\vec\Omega) \, + \, \varepsilon_i 
   \qquad (i=0,\ldots,3) \; ,
\end{equation}
where $S_{\! i}(\nu,\vec\Omega)$ are the Stokes parameters of the radiation
flowing through point $\vec x$ in the direction $\vec\Omega$, defined with
respect to the unit vectors $\vec{e}_a(\vec\Omega)$, $\vec{e}_b(\vec\Omega)$ 
of Fig.~1, $K_{ij}^{A}$ is the absorption matrix, and $\varepsilon_i$ 
are the emission coefficients in the four Stokes parameters.
The explicit expressions of the radiative transfer coefficients for the case 
we are concerned with can be derived by a simple generalization of the results 
contained in Section~14.2 of LL04. 
Since we have assumed that the lower level is unpolarized, and that the Zeeman 
splitting is negligible with respect to $\Delta \nu_{\rm D}$, the absorption 
matrix $K_{ij}^{\rm A}$ is proportional to the identity matrix, i.e., it is of 
the form
\begin{equation}
	K_{ij}^{\rm A} = \eta^{\, \rm A}_0(\nu, \vec{\Omega}) \, \delta_{ij} 
	\; .
\end{equation}
Due to the Doppler effect, an atom with velocity $\boldsymbol{\varv}$ absorbs 
radiation propagating in direction $\vec{\Omega}$ only at the frequency 
$\nu_0(1+\boldsymbol{\varv} \cdot \vec{\Omega}/c)$ (we assumed that the upper 
and lower levels are infinitely sharp). 
The absorption coefficient $\eta^{\rm A}_0(\nu, \vec{\Omega})$ is thus given by
\begin{equation}
	\eta_0^{\, \rm A}(\nu, \vec{\Omega}) = k_{\rm L}^{\, \rm A}(\vec{x}) 
	\, p(\nu, \vec{\Omega}) \; ,
\label{Eq:abs_coef}
\end{equation}
with
\begin{equation}
	p(\nu, \vec{\Omega}) = \int {\rm d}^3 \boldsymbol{\varv} \, 
	f(\boldsymbol{\varv}) \, \delta \left( \nu_0 + \nu_0 
	\frac{\boldsymbol{\varv} \cdot \vec{\Omega}}{c} - \nu \right) \; ,
\label{Eq:abs_prof}
\end{equation}
where $\delta$ is the Dirac-delta. 
The quantity $k_{\rm L}^{\, \rm A}(\vec{x})$ is the frequency-integrated 
absorption coefficient of the line, given by
\begin{equation}
	k_{\rm L}^{\, \rm A}(\vec{x}) = \frac{h \nu_0}{4 \pi} \, 
	{\mathcal N}_{\ell}(\vec{x}) \; 
	B(\alpha_\ell J_\ell \rightarrow \alpha_u J_u) \; ,
\end{equation}
with ${\mathcal N}_{\ell}(\vec{x})$ the number density of atoms in the lower 
level at point $\vec{x}$.

The expression for the emission coefficient in the four Stokes parameters
is obtained from Eq.~(7.16e) of LL04 (where the velocity-independent density 
matrix has now to be substituted with the velocity-space density matrix, and 
where, consistently with our assumptions, the profile $\phi(\nu_0-\nu)$ is now 
a Dirac-delta).
Considering that, due to the Doppler effect, the atoms having velocity 
$\boldsymbol{\varv}$ emit, along the direction $\vec \Omega$, at the frequency 
$\nu_0 (1 + \boldsymbol{\varv} \cdot \vec \Omega /c)$, we have 
\begin{equation}
	\varepsilon_i(\nu,\vec \Omega) = k_{\rm L}^{\, \rm A}(\vec{x}) 
	\int {\rm d}^3 \boldsymbol{\varv} \; f(\boldsymbol{\varv}) \; 
	\delta \left(\nu_0 + \nu_0 \, 
	\frac{\boldsymbol{\varv} \cdot \vec \Omega}{c}-\nu \right) \; 
	\sum_{KQ} \; w^{\, (K)}_{J_u J_\ell} \; 
	{\mathcal T}^K_{\,Q}(i,\vec\Omega) \;
	{\mathcal S}^K_Q(\boldsymbol{\varv}, \vec x) \; .
\label{Eq:emis_coef}
\end{equation}
Obviously, the preceding expressions imply that the only contribution to the
opacity and emissivity of the medium comes from transitions between the two
levels of the model atom. The case where a source of continuum opacity (and
emissivity) is also present is formally more complicated and will not be
treated here.

The radiative transfer equation can be formally solved. 
\begin{figure}[!t]
\centering
\includegraphics[width=0.5\textwidth]{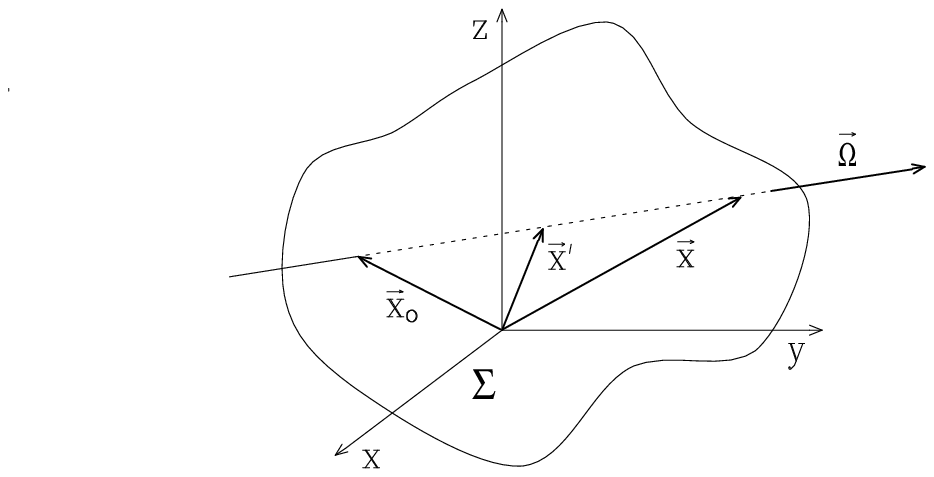}
\caption{A ray with direction $\,\vec\Omega\,$ enters the medium at point
$\,\vec x_0\,$, where its Stokes parameters are 
$S_{\! i}^{\rm (b)}(\nu,\vec\Omega)$. At point $\vec x$, the Stokes parameters 
are given by Eq.~(\ref{Eq:RT_formal-sol}).}
\end{figure}
Referring to Fig.~2, and using Eq.~(8.18) of LL04, the Stokes parameters 
at point $\vec x$ of the radiation at frequency $\nu$, flowing along the 
direction $\vec\Omega$ can be expressed in the form
\begin{equation}
   S_{\! i}(\nu,\vec\Omega) = \int_{\vec x_0}^{\vec x} {\rm d}s' 
   \int {\rm d}^3 \boldsymbol{\varv} \; f(\boldsymbol{\varv}) \; 
   \delta \left( \nu_0 + \nu_0 \, 
   \frac{\boldsymbol{\varv} \cdot \vec \Omega}{c} -\nu \right)
   k_{\rm L}^{\, \rm A}(\vec x^{\,\prime}) \;
   {\rm e}^{-\tau_\nu(\vec x,\vec x')}
   \sum_{KQ}\; w^{\, (K)}_{J_u J_\ell} \; {\mathcal T}^K_{\,Q}(i,\vec\Omega)
   \; {\mathcal S}^K_Q(\boldsymbol{\varv} , \vec x^{\,\prime}) 
   + \, {\rm e}^{-\tau_\nu(\vec x,\vec x_0)} \; 
   S_{\! i}^{\rm (b)}(\nu,\vec\Omega) \; , 
\label{Eq:RT_formal-sol}
\end{equation}
where $S_{\! i}^{\rm (b)}(\nu,\vec\Omega)$ is the Stokes vector of the radiation
entering the medium at point $\vec x_0$ along the direction $\vec\Omega$, $s'$
is the coordinate of $\vec x^{\,\prime}$ reckoned along $\vec\Omega$
($s'=|\vec x^{\,\prime}-\vec x_0|$), and $\tau_\nu(\vec x,\vec x^{\,\prime})$
is the optical depth at frequency $\nu$ between points $\vec x$ and
$\vec x^{\,\prime}$,
\begin{equation}
   \tau_\nu(\vec x,\vec x^{\,\prime}) =
   \int_{\vec x'}^{\vec x} \! {\rm d} s^{\prime \prime} 
   \eta^{\, \rm A}_0(\nu,\vec{\Omega}) =
   \int_{\vec x'}^{\vec x} \! {\rm d} s^{\prime \prime} 
   k_{\rm L}^{\, \rm A}(\vec x^{\,\prime\prime}) 
   \int {\rm d}^3 \boldsymbol{\varv} \, f(\boldsymbol{\varv}) \, 
   \delta \left( \nu_0 + \nu_0 \frac{\boldsymbol{\varv} \cdot \vec{\Omega}}{c} 
   - \nu \right) \; .
\label{Eq:opt_depth}
\end{equation}
It is now possible to find the expression for the radiation field tensor at 
point $\vec x$. Substituting Eq.~(\ref{Eq:RT_formal-sol}) into 
Eq.~(\ref{Eq:JKQ_cf}), we obtain two contributions and we can thus write
\begin{equation}
   \left[ J^K_Q(\nu_0) \right]_{\rm c.f.} = \left[ J^K_Q(\nu_0) \right]_{\rm I} 
   \, + \, \left[ J^K_Q(\nu_0) \right]_{\rm E} \; ,
\label{Eq:JKQ_IE}
\end{equation}
where the `internal' part $\left[ J^K_Q(\nu_0) \right]_{\rm I}$ is given by
\begin{equation}
  \left[ J^K_Q(\nu_0) \right]_{\rm I} = \oint \frac{{\rm d}\Omega}{4\pi} \; 
  \sum_{i=0}^3 \; {\mathcal T}^K_{\,Q}(i,\vec\Omega) 
  \int_{\vec x_0}^{\vec x} \! {\rm d} s' \int {\rm d}^3 
  \boldsymbol{\varv}^{\, \prime} f(\boldsymbol{\varv}^{\, \prime}) \, \delta 
  \left( \nu_0 \frac{\boldsymbol{\varv}^{\, \prime} - \boldsymbol{\varv}}{c} 
  \cdot \vec \Omega \right) \; k_{\rm L}^{\, \rm A}(\vec x^{\,\prime}) \;\,
  {\rm e}^{-\tau_{\nu^\prime}(\vec x,\vec x')} \; \sum_{K'Q'} \; 
  w^{\, (K')}_{J_u J_\ell} \; {\mathcal T}^{K'}_{\,Q'}(i,\vec\Omega) \;
  {\mathcal S}^{K'}_{Q'}(\boldsymbol{\varv}^{\, \prime},\vec x^{\,\prime}) \; ,
\label{Eq:JKQ_I1}
\end{equation}
where
\begin{equation}
	\nu^{\prime} = \nu_0 + \nu_0 \, 
	\frac{\boldsymbol{\varv} \cdot \vec\Omega}{c} \; , 
\label{Eq:nup}
\end{equation}
and the `external' part $\left[ J^K_Q(\nu_0) \right]_{\rm E}$, originating 
from the boundary conditions, by
\begin{equation}
  \left[ J^K_Q(\nu_0) \right]_{\rm E} =
  \oint \frac{{\rm d}\Omega}{4\pi} \; \sum_{i=0}^3 \; 
  {\mathcal T}^K_{\,Q}(i,\vec\Omega) \;
  {\rm e}^{-\tau_{\nu^\prime}(\vec x,\vec x_0)} \; 
  S_{\! i}^{\rm (b)}(\nu^{\prime},\vec\Omega) \; .
\end{equation}
Equation~(\ref{Eq:JKQ_I1}) can be cast in a simpler form by changing the double 
integral in ${\rm d}\Omega$ and ${\rm d} s'$ into a volume integral. 
Since
\begin{equation}
   {\rm d}^3 \vec x^{\,\prime} = 
   (\vec x-\vec x^{\,\prime})^2 \; {\rm d}\Omega \, {\rm d} s' \; ,
\end{equation}
we get
\begin{equation}
   \left[ J^K_Q(\nu_0) \right]_{\rm I} = \int \! {\rm d}^3 
   \vec x^{\,\prime} \; \frac{k_{\rm L}^{\, \rm A}(\vec x^{\,\prime}) \;
   {\rm e}^{-\tau_{\nu^\prime}(\vec x,\vec x')}}
   {4\pi(\vec x-\vec x^{\,\prime})^2}
   \int {\rm d}^3 \boldsymbol{\varv}^{\, \prime} \, 
   f(\boldsymbol{\varv}^{\, \prime}) \,
   \delta \left( \nu_0 \frac{\boldsymbol{\varv}^{\, \prime} - 
   \boldsymbol{\varv}}{c} \cdot \vec\Omega \right) \sum_{i=0}^3 \; 
   {\mathcal T}^K_{\,Q}(i,\vec\Omega) \sum_{K'Q'} \; 
   w^{\, (K')}_{J_u J_\ell} \; {\mathcal T}^{K'}_{\,Q'}(i,\vec\Omega) \; 
   {\mathcal S}^{K'}_{Q'}(\boldsymbol{\varv}^{\, \prime}, 
   \vec x^{\,\prime}) \; .
\label{Eq:JKQ_I}
\end{equation}

\section{Coupled equations for the velocity dependent irreducible components of 
the source function}
We can now substitute the expression of the radiation field tensor at point 
$\vec x$ into the statistical equilibrium equation. 
From Eqs.~(\ref{Eq:SF_SEE}), (\ref{Eq:JKQ_IE}), and (\ref{Eq:JKQ_I}) we obtain
\begin{equation}
\begin{split}
	\left[ 1+ \epsilon+ \delta_{\,u}^{\, (K)}(\boldsymbol{\varv}) \right]
	{\mathcal S}^K_Q(\boldsymbol{\varv}, \vec x\,) + {\rm i} \, H_u 
   	\sum_{Q'} \, {\mathcal K}^K_{QQ'} \, {\mathcal S}^K_{Q'} &
   	(\boldsymbol{\varv}, \vec x \,) =
   	\delta_{K0} \, \delta_{Q0} \; \epsilon \, B_T(\nu_0) + 
   	w^{\, (K)}_{J_u J_\ell} \, (-1)^Q \left[ J^K_{-Q}(\nu_0) 
	\right]_{\rm E}+ \\
   	& + \int \! {\rm d}^3 \vec x^{\,\prime} \,
   	\frac{k_{\rm L}^{\, \rm A}(\vec x^{\,\prime})}
   	{4\pi(\vec x-\vec x^{\,\prime})^2} 
   	\int \! {\rm d}^3 \boldsymbol{\varv}^{\, \prime} \, 
	f(\boldsymbol{\varv}^{\, \prime}) \! \sum_{K'Q'} 
	G_{KQ,K'Q'}(\boldsymbol{\varv}, \vec x \, ; 
	\boldsymbol{\varv}^{\, \prime}, \vec x^{\,\prime}) \;
   	{\mathcal S}^{K'}_{Q'}(\boldsymbol{\varv}^{\, \prime},
	\vec x^{\,\prime}) \; ,
\label{Eq:Coupled_SF1}
\end{split}
\end{equation}
where
\begin{equation}
	G_{KQ,K'Q'}(\boldsymbol{\varv}, \vec x \, ; 
	\boldsymbol{\varv}^{\, \prime},\vec x^{\,\prime})
   	= {\rm e}^{-\tau_{\nu^\prime}(\vec x,\vec x')} \, 
   	\delta \left( \nu_0 \frac{\boldsymbol{\varv}^{\, \prime} - 
	\boldsymbol{\varv}}{c} \cdot \vec \Omega \right) \, 
	w^{\, (K)}_{J_u J_\ell} \; w^{\, (K')}_{J_u J_\ell} 
   	\, \sum_{i=0}^3 \; (-1)^Q \; {\mathcal T}^K_{-Q}(i,\vec\Omega) \,
   	{\mathcal T}^{K'}_{\,Q'}(i,\vec\Omega) \; .
\label{Eq:GKQ}
\end{equation}
The quantities $G_{KQ,K'Q'}(\boldsymbol{\varv}, \vec x \, ; 
\boldsymbol{\varv}^{\,\prime}, \vec x^{\,\prime})$ appearing in this equation 
represent a factor (having the dimension of the inverse of a frequency) which 
weights the amount of coupling between the statistical tensor $\rho^K_Q$ of the 
atoms having velocity $\boldsymbol{\varv}$ at point $\vec x$ and the 
statistical tensor $\rho^{K'}_{Q'}$ of the atoms having velocity 
$\boldsymbol{\varv}^{\, \prime}$ at point $\vec x^{\,\prime}$. 
They are a generalization of similar quantities introduced in LL04 and can be 
referred to as {\it velocity dependent multipole coupling coefficients}.
The Dirac-delta appearing in their expression is responsible for the fact that 
two such multipoles can be coupled only if the velocity difference, 
$\boldsymbol{\varv} - \boldsymbol{\varv}^{\,\prime}$ is perpendicular to the 
unit vector $\vec \Omega$ which specifies the direction 
$\vec x - \vec x^{\,\prime}$.

Equation~(\ref{Eq:Coupled_SF1}) is a system of linear, non-homogeneous, 
integral equations in the unknowns ${\mathcal S}^K_Q(\boldsymbol{\varv}, 
\vec x\,)$, the velocity dependent irreducible components of the source 
function, which can in principle be solved once the properties of the medium 
and the boundary conditions are specified. 
When the values of these components are known at each point, the Stokes 
parameters of the radiation emerging from the medium can be computed by 
applying Eq.~(\ref{Eq:RT_formal-sol}). 
It should be remarked that, owing to a property of the tensor 
${\mathcal T}^K_Q$, whose proof can be found in App.~20 of LL04, it can be 
shown that Eq.~(\ref{Eq:Coupled_SF1}) decouples in two different sets of 
equations involving, respectively, the components with $K=0,2$ and those with 
$K=1$. 
In the latter set, the only source term is $[ J^1_{-Q}(\nu_0)]_{\rm E}$, 
which vanishes unless the boundary radiation field has some contribution 
arising from circular polarization. 
Excluding this case of limited interest, all the components 
${\mathcal S}^1_Q(\boldsymbol{\varv}, \vec x\,)$ are everywhere zero in the 
medium.

For further developments it is however more practical to rewrite 
Eq.~(\ref{Eq:Coupled_SF1}) in an alternative form by substitution of 
Eq.~(\ref{Eq:GKQ}) and by introducing the compact symbol 
$\Gamma_{K Q, K' Q'}(\vec \Omega)$, whose main properties are collected in 
App.~20 of LL04. 
Its definition, that we recall here for completeness, is the following  
\begin{equation}
   \Gamma_{K Q, K' Q'} (\vec \Omega) = \sum_{i=0}^3 \; (-1)^Q \; 
   {\mathcal T}^K_{-Q}(i,\vec\Omega) \, {\mathcal T}^{K'}_{\,Q'}(i,\vec\Omega)
   \; .
\end{equation}
With these transformations we obtain
\begin{equation}
\begin{split}
	\left[ 1+ \epsilon+ \delta_{\,u}^{\, (K)}(\boldsymbol{\varv}) 
	\right] & \, {\mathcal S}^K_Q(\boldsymbol{\varv}, \vec x) \, + \, 
	{\rm i} \, H_u \sum_{Q'} \, {\mathcal K}^K_{QQ'} \, 
	{\mathcal S}^K_{Q'}(\boldsymbol{\varv}, \vec x) = 
   	\delta_{K0} \, \delta_{Q0} \, \epsilon \, B_T(\nu_0) \, + \, 
   	w^{\, (K)}_{J_u J_\ell} \, (-1)^Q 
	\left[ J^K_{-Q}(\nu_0) \right]_{\rm E} + \\
   	& + \int \! {\rm d}^3 \vec x^{\,\prime} \,
   	\frac{k_{\rm L}^{\, \rm A}(\vec x^{\,\prime})}
   	{4\pi(\vec x-\vec x^{\,\prime})^2} \, 
   	{\rm e}^{-\tau_{\nu^\prime}(\vec x,\vec x')} \,
   	\int \! {\rm d}^3 \boldsymbol{\varv}^{\, \prime} \, 
	f(\boldsymbol{\varv}^{\, \prime}) \, \delta \left( \nu_0 
	\frac{\boldsymbol{\varv}^{\, \prime} - \boldsymbol{\varv}}{c} \cdot 
	\vec \Omega \right) \sum_{K'Q'} 
   	w^{\, (K)}_{J_u J_\ell} \; w^{\, (K')}_{J_u J_\ell} \, 
   	\Gamma_{K Q, K' Q'} (\vec \Omega) \, 
   	{\mathcal S}^{K'}_{Q'}(\boldsymbol{\varv}^{\, \prime}, 
	\vec x^{\,\prime}) \; .
\label{Eq:Coupled_SF2}
\end{split}
\end{equation}
This is a very general set of coupled equations for the velocity-dependent 
source function that in principle can be solved numerically by suitable 
discretization of $\mathcal{S}^K_Q(\boldsymbol{\varv},\vec{x})$ over the 
velocity-space and the physical-space.

We now assume that the velocity distribution $f(\boldsymbol{\varv})$ is a 
Maxwellian characterized by the thermal velocity $\varv_t$ (possibly containing 
the contribution of microturbulent velocities)
\begin{equation}
	f(\boldsymbol{\varv}) = \frac{1}{\varv_t^3 \, \pi^{3/2}} 
	\, {\rm e}^{-\varv^2/ \varv_t^2} \; .
\end{equation}
For fixed $\vec x$ and $\vec x^\prime$ (which implies a fixed direction 
$\vec \Omega$ joining $\vec x^\prime$ with $\vec x$), and for a fixed 
velocity $\boldsymbol{\varv}$, we introduce a right-handed, Cartesian 
coordinate system $(\vec e_{\rm a}, \vec e_{\rm b}, \vec e_{\rm c})$ in the 
velocity space. 
The unit vector $\vec e_{\rm c}$ is directed along $\vec \Omega$ (coinciding 
with it), while the other two vectors are perpendicular to $\vec \Omega$ and 
for the rest arbitrary.
In this system we obviously have
\begin{equation}
	\boldsymbol{\varv} = \varv_{\rm a} \, \vec e_{\rm a} + \varv_{\rm b} \, 
   	\vec e_{\rm b} + \varv_{\rm c} \, \vec e_{\rm c}  \; , 
   	\qquad \boldsymbol{\varv}^{\, \prime} = \varv^\prime_{\rm a} \, 
	\vec e_{\rm a} + \varv^\prime_{\rm b} \, \vec e_{\rm b} + 
	\varv^\prime_{\rm c} \, \vec e_{\rm c}  \; .
\end{equation}
With these position, we can perform the following formal substitution
in the integral in ${\rm d}^3 \boldsymbol{\varv}^{\, \prime}$
\begin{equation}
	\int {\rm d}^3 \boldsymbol{\varv}^{\, \prime} \, 
	f(\boldsymbol{\varv}^{\, \prime}) \, \delta \left( \nu_0 
	\frac{\boldsymbol{\varv}^{\, \prime} - \boldsymbol{\varv}}{c} 
	\cdot \vec \Omega \right) \rightarrow 
   	\int \frac{{\rm d} \varv^\prime_{\rm a}}{\varv_{\rm t}} 
   	\int \frac{{\rm d} \varv^\prime_{\rm b}}{\varv_{\rm t}}
   	\int \frac{{\rm d} \varv^\prime_{\rm c}}{\varv_{\rm t}}
   	\; \frac{{\rm e}^{-(\varv_{\rm a}^{\prime \, 2} + 
   	\varv_{\rm b}^{\prime \, 2} + \varv_{\rm c}^{\prime \, 2})/ 
   	\varv_{\rm t}^2}}{\pi^{3/2}} \;
   	\delta \left( \nu_0 \frac{\varv_{\rm c}^{\prime} - \varv_{\rm c}}{c} 
	\right) \; .
\end{equation}
Due to the presence of the Dirac's delta, the integral in 
${\rm d} \varv_{\rm c}^{\prime}$ is immediately performed, and the last term 
in the right-hand side of Eq.~(\ref{Eq:Coupled_SF2}) acquires the form 
\begin{equation}
   \int \! {\rm d}^3 \vec x^{\,\prime} \,
   \frac{k_{\rm L}^{\, \rm A}(\vec x^{\,\prime})}
   {4\pi(\vec x-\vec x^{\,\prime})^2} \; 
   {\rm e}^{-\tau_{\nu^{\prime}}(\vec x,\vec x')} \,
   \int \frac{{\rm d}\varv_{\rm a}^{\prime}}{\varv_{\rm t}}
   \int \frac{{\rm d}\varv_{\rm b}^{\prime}}{\varv_{\rm t}} \;
   \frac{{\rm e}^{-(\varv_{\rm a}^{\prime \, 2}
   +\varv_{\rm b}^{\prime \, 2} + \varv_{\rm c}^2)/\varv_{\rm t}^2}}{\pi^{3/2}} 
   \frac{1}{\Delta \nu_{\rm D}} \sum_{K'Q'} w^{\, (K)}_{J_u J_\ell} \; 
   w^{\, (K')}_{J_u J_\ell} \, \Gamma_{K Q, K' Q'} (\vec \Omega) \,
   {\mathcal S}^{K'}_{Q'}(\boldsymbol{\varv}^{\, \prime}_{\ast}, 
   \vec x^{\,\prime}) \; ,
\end{equation}
where the vector $\boldsymbol{\varv}^{\, \prime}_{\ast}$ in the argument of 
${\mathcal S}^{K'}_{Q'}$ has components $(\varv^\prime_{\rm a}, 
\varv^\prime_{\rm b}, \varv_{\rm c})$, and where the Doppler width 
$\Delta \nu_{\rm D}$ is given by
\begin{equation}
	\Delta \nu_{\rm D} = \nu_0 \frac{\varv_t}{c} \; .
\end{equation}
We also observe that under the assumption that the velocity distribution is a 
Maxwellian characterized by the thermal velocity $\varv_t$, the absorption 
coefficient (see Eq.~(\ref{Eq:abs_coef})) does not depend any longer on the 
propagation direction of the radiation ($\vec{\Omega}$), and it is given by
\begin{equation}
	\eta_0^{\, \rm A}(\nu) = k_{\rm L}^{\, \rm A}(\vec{x}) 
	\, p_{\rm M}(\nu - \nu_0) \; ,
\end{equation}
with
\begin{equation}
	p_{\rm M}(\nu- \nu_0) = \frac{1}{\sqrt{\pi} \, \Delta \nu_{\rm D}} 
	\, {\rm e}^{-(\nu - \nu_0)^2/\Delta \nu_{\rm D}^2} \; .
\end{equation}
Under the same assumption, the optical depth $\tau_{\nu}(\vec{x}, 
\vec{x}^{\, \prime})$ (see Eq.~(\ref{Eq:opt_depth})) is given by
\begin{equation}
	\tau_\nu(\vec x,\vec x^{\,\prime}) = p_{\rm M}(\nu - \nu_0)
	\int_{\vec x'}^{\vec x} \! {\rm d} s^{\prime \prime} 
	k_{\rm L}^{\, \rm A}(\vec x^{\,\prime\prime}) \; .
\end{equation}

\section{Equations for a plane-parallel semi-infinite stellar atmosphere}
We now consider the particular case of a plane-parallel, semi-infinite stellar 
atmosphere. 
In this case all the physical quantities of the medium depend on a single 
coordinate, the height in the atmosphere, that we assume as the $z$-axis of our 
fixed reference system of Fig.~1. 
As a consequence, the irreducible components of the velocity-dependent source 
function only vary with the height $z$. 
We keep assuming that the velocity distribution is Maxwellian and we introduce 
the line optical depth $t_{\rm L}$ through the equation 
\begin{equation}
	{\rm d} t_{\rm L} = -\frac{k_{\rm L}^{\, \rm A}(z)}
	{\Delta \nu_{\rm D}} \, {\rm d} z \; .
\end{equation}
Assuming that the stellar atmosphere is not illuminated by external sources
of radiation, Eq.~(\ref{Eq:Coupled_SF2}) takes the form
\begin{equation}
\begin{split}
	\Big[ 1 \, + & \, \epsilon + \delta_{\,u}^{\, (K)}(\boldsymbol{\varv} 
	\,) \Big] \, {\mathcal S}^K_Q(\boldsymbol{\varv}, t_{\rm L}) + 
	{\rm i} \, H_u \sum_{Q'} \, {\mathcal K}^K_{QQ'} \, 
	{\mathcal S}^K_{Q'}(\boldsymbol{\varv}, t_{\rm L}) = \,
   	\delta_{K0} \, \delta_{Q0} \; \epsilon \, B_T(\nu_0) \, + \\
   	& +\int_0^\infty \! {\rm d} t'_{\rm L} \! \int_{-\infty}^{\infty} \!\!
   	{\rm d} x' \! \int_{-\infty}^{\infty} \!\! {\rm d} y' \,
   	\frac{1}{4\pi(\vec x - \vec x^{\,\prime})^2} \,
   	{\rm e}^{-\tau_{\nu^\prime}(\vec x,\vec x')} \,
   	\int \frac{{\rm d}\varv_{\rm a}^\prime}{\varv_{\rm t}}
   	\int \frac{{\rm d}\varv_{\rm b}^\prime}{\varv_{\rm t}} \;
   	\frac{{\rm e}^{-(\varv_{\rm a}^{\prime \, 2}
   	+\varv_{\rm b}^{\prime \, 2} + 
	\varv_{\rm c}^2)/\varv_{\rm t}^2}}{\pi^{3/2}} 
   	\sum_{K'Q'} w^{\, (K)}_{J_u J_\ell} \; w^{\, (K')}_{J_u J_\ell} \, 
   	\Gamma_{K Q, K' Q'} (\vec \Omega) \, 
   	{\mathcal S}^{K'}_{Q'}(\boldsymbol{\varv}^{\, \prime}_{\ast},
	t'_{\rm L}) \;\; , 
\label{Eq:Coupled_SF_PP1}
\end{split}
\end{equation}
where the indices $K$ and $K'$ are restricted to the values 0 and 2. 

\begin{figure}[!t]
\centering
\includegraphics[width=0.5\textwidth]{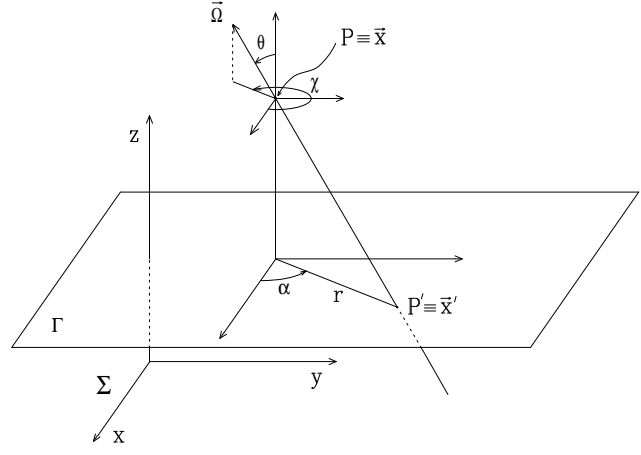}
\caption{In the reference system $\Sigma$ we consider a fixed point {\rm P}
located at height $z$ corresponding to line optical depth $t_{\rm L}$. 
The point ${\rm P}'$ lies on the plane $\Gamma$ parallel to the \hbox{$x$-$y$}
plane; its height is $z'$ corresponding to line optical depth $t'_{\rm L}$. 
The cylindrical coordinates $(r,\alpha)$ define the position of ${\rm P}'$ in 
the plane $\Gamma$. 
The angles $\theta$ and $\chi$ specify the direction $\vec\Omega$.}
\end{figure}
The integral over $x'$ and $y'$ can be performed by introducing cylindrical
coordinates and following a procedure similar to the one developed in App.~21
of LL04. Referring to the geometry of Fig.~3 (which represents the case 
$t'_{\rm L}>t_{\rm L}$ or $z'<z$), and introducing the cylindrical coordinates
$r$ and $\alpha$ of the point ${\rm P}'$, one has
\begin{equation}
   {\rm d} x' \, {\rm d} y' = r \; {\rm d} r \, {\rm d} \alpha \; .
\end{equation}
On the other hand
\begin{equation}
   (\vec x-\vec x^{\,\prime})^2=(z-z')^2 + r^2 \; ,
\end{equation}
and
\begin{equation}
   r = (z-z') \tan \theta \; , \qquad
   {\rm d} r = \frac{z-z'}{\cos^2 \! \theta} \, {\rm d} \theta \; .
\end{equation}
From these relations we get
\begin{equation}
   \frac{{\rm d} x' \, {\rm d} y'}{(\vec x - \vec x^{\,\prime})^2} \, = \, 
   \tan \theta \; {\rm d} \theta \, {\rm d} \alpha \; ,
\end{equation}
and since $\chi=\alpha+\pi$, the double integral over $x'$ and $y'$ can be
transformed into an integral over the angles $\theta$ and $\chi$ specifying the
direction $\vec\Omega$. In Eq.~(\ref{Eq:Coupled_SF_PP1}) one can then perform 
the formal substitution
\begin{equation}
   \int_{-\infty}^{\infty} \!\! {\rm d} x' \!
   \int_{-\infty}^{\infty} \!\! {\rm d} y' \,
   \frac{1}{4\pi(\vec x-\vec x^{\,\prime})^2} \rightarrow 
   \frac{1}{4 \pi} \int_0^{2 \pi} \! {\rm d} \chi \int_0^{\pi/2} \! 
   {\rm d} \theta \;\, \tan\theta \; .
\end{equation}
Moreover, introducing the reduced frequency distance from line center through
the usual expression
\begin{equation}
   \xi = \frac{\nu - \nu_0}{\Delta \nu_{\rm D}} \; ,
\end{equation}
and the normalized profile
\begin{equation}
   \varphi(\xi) = \frac{1}{\sqrt{\pi}} \, {\rm e}^{-\xi^2} \; ,
\label{Eq:abs_prof_red}
\end{equation}
one has
\begin{equation}
   \tau_{\nu'}(\vec x, \vec x^{\, \prime}) = \frac{(t'_{\rm L} - t_{\rm L}) \, 
   \varphi(\xi')}{\cos \theta} \; ,
\end{equation}
with (recalling Eq.~(\ref{Eq:nup}))
\begin{equation}
   \xi' = \frac{\nu' - \nu_0}{\Delta \nu_{\rm D}} = 
   \frac{\nu_0 \, \varv_{\rm c}}{\Delta \nu_{\rm D} \, c}  \; . 
\end{equation}
Taking into account these transformations, and performing a similar analysis
for the case $t'_{\rm L} < t_{\rm L}$, Eq.~(\ref{Eq:Coupled_SF_PP1}) becomes
\begin{equation}
\begin{split}
	\Big[ 1 & \, + \epsilon+ \delta_{\,u}^{\, (K)}(\boldsymbol{\varv} \,) 
	\Big] {\mathcal S}^K_Q(\boldsymbol{\varv}, t_{\rm L}) + {\rm i} \, 
	H_u \sum_{Q'} \, {\mathcal K}^K_{QQ'} \, 
	{\mathcal S}^K_{Q'}(\boldsymbol{\varv}, t_{\rm L}) = \,
   	\delta_{K0} \, \delta_{Q0} \; \epsilon \, B_T(\nu_0) + \\
   	& + \int_0^{\infty} \! {\rm d} t'_{\rm L} \; \frac{1}{4 \pi} 
   	\int_0^{2 \pi} \! {\rm d} \chi \int_{\theta_1}^{\theta_2} \! 
   	{\rm d} \theta \, |\tan\theta \,| \; {\rm e}^{- (t'_{\rm L} - 
	t_{\rm L}) \, \varphi(\xi') / \cos \theta} \,
   	\int \frac{{\rm d}\varv_{\rm a}^\prime}{\varv_{\rm t}}
   	\int \frac{{\rm d}\varv_{\rm b}^\prime}{\varv_{\rm t}} \;
   	\frac{{\rm e}^{-(\varv_{\rm a}^{\prime \, 2} + 
	\varv_{\rm b}^{\prime \, 2} + 
	\varv_{\rm c}^2)/\varv_{\rm t}^2}}{\pi^{3/2}}
   	\sum_{K'Q'} w^{\, (K)}_{J_u J_\ell} \; w^{\, (K')}_{J_u J_\ell} \, 
   	\Gamma_{K Q, K' Q'} (\vec \Omega) \, 
   	S^{K'}_{Q'}(\boldsymbol{\varv}^{\, \prime}_{\ast}, t'_{\rm L}) \; ,
\end{split}
\end{equation}
the interval $(\theta_1,\theta_2)$ being $(0,\pi /2)$ if 
$t'_{\rm L} > t_{\rm L}$ and $(\pi /2, \pi)$ if $t'_{\rm L} < t_{\rm L}$.

It is now necessary to specify the unit vectors $\vec e_{\rm a}$ and 
$\vec e_{\rm b}$ that have been left undefined. For a given direction 
$\vec \Omega$  we set
\begin{align}
   \vec e_{\rm a} & = \vec e_\theta = \cos \theta \, \cos \chi \, 
   \vec i  + \cos \theta \, \sin \chi \, \vec j -\sin \theta \, 
   \vec k \;\; , \nonumber \\
   \vec e_{\rm b} & = \vec e_\chi = -\sin \chi \, \vec i 
   + \cos \chi \, \vec j \; , \\
   \vec e_{\rm c} & = \vec e_r = \vec \Omega = \sin \theta \, 
   \cos \chi \, \vec i  + \sin \theta \, \sin \chi \, \vec j 
   +\cos \theta \, \vec k \nonumber \; .
\end{align}
Given the velocity components $\varv'_{\rm a}$ and $\varv'_{\rm b}$, the 
velocity $\boldsymbol{\varv}^{\, \prime}_{\ast}$ is thus given by
\begin{equation}
\begin{split}
	\boldsymbol{\varv}^{\, \prime}_{\ast} = \varv'_{\rm a} \, 
	\vec e_{\rm a} + \varv'_{\rm b} \, \vec e_{\rm b} + \varv_{\rm c} \, 
	\vec e_{\rm c} = & \, \left[ (\cos \theta \, \varv'_{\rm a} + 
	\sin \theta \, \varv_{\rm c}) \cos \chi - \sin \chi \, 
	\varv'_{\rm b} \right] \, \vec i \nonumber \\
   	& +\left[ (\cos \theta \, \varv'_{\rm a} + \sin \theta \, 
	\varv_{\rm c}) \sin \chi + \cos \chi \, \varv'_{\rm b} \right] \, 
	\vec j \\
   	& -\left[ \sin \theta \, \varv'_{\rm a} - \cos \theta \, \varv_{\rm c} 
   	\right] \, \vec k \; .
\end{split}
\end{equation}
This equation allows us to find the modulus, $\varv'_{\ast}$, polar angle, 
$\theta_{\ast}$, and azimuth, $\chi_{\ast}$, of the velocity 
$\boldsymbol{\varv}^{\, \prime}_{\ast}$ in terms of $\varv'_{\rm a}$, 
$\varv'_{\rm b}$, and $\varv_{\rm c}$. One gets
\begin{align}
   \varv'_{\ast} & = \sqrt{\varv_{\rm a}^{\prime 2} + 
   \varv_{\rm b}^{\prime 2} + \varv_{\rm c}^2} \; , \nonumber \\
   \varv'_{\ast} \, \cos \theta_{\ast} & = -\sin \theta \, 
   \varv'_{\rm a} + \cos \theta \, \varv_{\rm c}  \; , \nonumber \\
   \varv'_{\ast} \, \sin \theta_{\ast} \, \cos \chi_{\ast} & =  
   (\cos \theta \, \varv'_{\rm a} + \sin \theta \, \varv_{\rm c}) \cos \chi
   -\sin \chi \, \varv'_{\rm b} \; , \\
   \varv'_{\ast} \, \sin \theta_{\ast} \, \sin \chi_{\ast} & =  
   (\cos \theta \, \varv'_{\rm a} + \sin \theta \, \varv_{\rm c}) \sin \chi
   + \cos \chi \, \varv'_{\rm b} \nonumber \; ,
\end{align}
the combination of the last two equations giving
\begin{equation}
   \varv'_{\ast} \, \sin \theta_{\ast} \, {\rm e}^{\, {\rm i} \, \chi_{\ast}} 
   = (\cos \theta \, \varv'_{\rm a} + \sin \theta \, \varv_{\rm c} + {\rm i} 
   \, \varv'_{\rm b}) \, {\rm e}^{\, {\rm i} \, \chi} \; ,
\end{equation}
or
\begin{equation}
   {\rm e}^{- {\rm i} \, \chi_{\ast}} = \frac{\cos \theta \, \varv'_{\rm a} + 
   \sin \theta \, \varv_{\rm c} - {\rm i} \, \varv'_{\rm b}} 
   {\sqrt{(\cos \theta \, \varv'_{\rm a} + \sin \theta \, \varv_{\rm c})^2 + 
   \varv_{\rm b}^{\prime 2}}} \; {\rm e}^{- {\rm i} \, \chi} \; .  
   \label{Eq:exp_chiast}
\end{equation}
The previous equations can also be inverted to give
\begin{align}
   \varv'_{\rm a} & = \varv'_{\ast} \, \sin \theta_{\ast} \, 
   \cos \theta \, \cos(\chi_{\ast} - \chi) - \varv'_{\ast} \, 
   \cos \theta_{\ast} \, \sin \theta \; , \nonumber \\
   \varv'_{\rm b} & = \varv'_{\ast} \, \sin \theta_{\ast} \, 
   \sin(\chi_{\ast} - \chi) \; , \\
   \varv_{\rm c} & = \varv'_{\ast} \, \sin \theta_{\ast} \, 
   \sin \theta \, \cos(\chi_{\ast} - \chi) + \varv'_{\ast} \, 
   \cos \theta_{\ast} \, \cos \theta \nonumber \; .
\end{align}
We now consider the simplified case of cylindrical symmetry. This implies
the absence of a deterministic magnetic field \citep[the case of a turbulent
magnetic feld can be handled with a slight modification of the formalism; 
see Sect. 14.2 of LL04, and Appendix~A of][]{JTB99}. 
In a cylindrically symmetric environment the velocity-dependent irreducible 
components of the source function, ${\mathcal S}^K_Q(\boldsymbol{\varv}, 
t_{\rm L})$, have a dependence on the azimuth $\chi_\varv$ of the velocity of 
the form 
\begin{equation}
	{\mathcal S}^K_Q(\boldsymbol{\varv}, t_{\rm L}) = 
   	{\mathcal S}^K_Q(\varv,\theta_\varv,\chi_\varv,t_{\rm L}) = 
   	{\mathcal S}^K_Q(\varv,\theta_\varv,0,t_{\rm L}) \, 
   	{\rm e}^{-{\rm i} \, Q \, \chi_\varv } \; ,
\end{equation}
where ${\mathcal S}^K_Q(\varv,\theta_\varv,0, t_{\rm L})$ is the value of the 
source function corresponding to $\chi_\varv=0$. 
Taking into account this property, we can limit ourselves to find the coupled 
equations for the quantities 
${\mathcal S}^K_Q(\varv,\theta_\varv,0,t_{\rm L})$, thus finding 
\begin{equation}
\begin{split}
	\left[ 1 + \epsilon+ \delta_{\,u}^{\, (K)}(\boldsymbol{\varv} \,) 
	\right] & \, {\mathcal S}^K_Q(\varv, \theta_\varv, 0, t_{\rm L}) = 
   	\delta_{K0} \, \delta_{Q0} \; \epsilon \, B_T(\nu_0) \\ 
   	& + \int_0^{\infty} \! {\rm d} t'_{\rm L} \; \frac{1}{4 \pi} 
   	\int_0^{2 \pi} \! {\rm d} \chi \int_{\theta_1}^{\theta_2} \! 
   	{\rm d} \theta \, |\tan\theta \,| \;
   	{\rm e}^{-(t'_{\rm L} - t_{\rm L}) \, \varphi(\xi') / \cos \theta} \,
   	\int \frac{{\rm d}\varv_{\rm a}^{\prime}}{\varv_{\rm t}}
   	\int \frac{{\rm d}\varv_{\rm b}^{\prime}}{\varv_{\rm t}} \;
   	\frac{{\rm e}^{-(\varv_{\rm a}^{\prime \, 2} + 
	\varv_{\rm b}^{\prime \, 2} 
   	+ \varv_{\rm c}^2)/\varv_{\rm t}^2}}{\pi^{3/2}} \\
   	& \qquad \qquad 
   	\times \sum_{K'Q'} w^{\, (K)}_{J_u J_\ell} \; w^{\, (K')}_{J_u J_\ell} 
	\, \Gamma_{K Q, K' Q'} (\vec \Omega) \, {\rm e}^{-{\rm i} \, Q' 
	\chi_{\ast}} \, {\mathcal S}^{K'}_{Q'} \! (\varv^{\, \prime}_{\ast}, 
	\theta_{\ast}, 0, t'_{\rm L}) \; . 
\label{Eq:Coupled_SF_PP2}
\end{split}
\end{equation}
where, according to Eq.~(\ref{Eq:exp_chiast})
\begin{equation}
   {\rm e}^{-{\rm i} \, Q' \chi_{\ast}} = \left[ \frac{\cos \theta \, 
   \varv'_{\rm a} + \sin \theta \, \varv_{\rm c} - {\rm i} \, \varv'_{\rm b}} 
   {\sqrt{(\cos \theta \, \varv'_{\rm a} + \sin \theta \, \varv_{\rm c})^2 + 
   \varv_{\rm b}^{\prime 2}}} \right]^{Q'} \, {\rm e}^{-{\rm i}\, Q' \chi} \; . 
   \label{Eq:exp_Qchiast}   
\end{equation}
We remind that the quantity $\varv_{\rm c}$ appearing in this equation is 
defined by
\begin{equation}
	\varv_{\rm c} = \boldsymbol{\varv} \cdot \vec \Omega = \varv \, 
	(\cos \theta_\varv \cos \theta + \sin \theta_\varv \sin \theta 
	\cos \chi) \; .
\end{equation}
From Eq.~(\ref{Eq:exp_Qchiast}), one can notice that the integrand in
Eq.~(\ref{Eq:Coupled_SF_PP2}) is an odd function of $\varv'_{\rm b}$. 
When integrating over $\varv'_{\rm b}$ from $-\infty$ to $\infty$, one can 
thus perform the following substitutions
\begin{align}
   \left[ \frac{\cos \theta \, \varv'_{\rm a} + \sin \theta \, \varv_{\rm c} - 
   {\rm i} \, \varv'_{\rm b}}{\sqrt{(\cos \theta \, \varv'_{\rm a} + 
   \sin \theta \, \varv_{\rm c})^2 + \varv_{\rm b}^{\prime 2}}} \right]^{\pm 1} 
   & \rightarrow 
   \frac{\cos \theta \, \varv'_{\rm a} + \sin \theta \, \varv_{\rm c}} 
   {\sqrt{(\cos \theta \, \varv'_{\rm a} + \sin \theta \, \varv_{\rm c})^2 + 
   \varv_{\rm b}^{\prime 2}}} \; , \nonumber \\  
   \left[ \frac{\cos \theta \, \varv'_{\rm a} + \sin \theta \, \varv_{\rm c} - 
   {\rm i} \, \varv'_{\rm b}}{\sqrt{(\cos \theta \, \varv'_{\rm a} + 
   \sin \theta \, \varv_{\rm c})^2 + \varv_{\rm b}^{\prime 2}}} \right]^{\pm 2} 
   & \rightarrow
   \frac{(\cos \theta \, \varv'_{\rm a} + \sin \theta \, \varv_{\rm c})^2  -  
   \varv_{\rm b}^{\prime 2}}{(\cos \theta \, \varv'_{\rm a} + \sin \theta \, 
   \varv_{\rm c})^2 + \varv_{\rm b}^{\prime 2}} \; .
\end{align}
Finally, one can notice that the structure of Eq.~(\ref{Eq:Coupled_SF_PP2}) is 
such to be consistent with our hypothesis on the behavior of the source 
function with the azimuth of the velocity. 
This is easily proven by considering the fact that the quantity 
$\Gamma_{K Q, K' Q'} (\vec \Omega)$ depends on the angle $\chi$ through an 
exponential of the form $\exp[{\rm i} (Q'-Q)\chi]$. 
    
It is convenient to perform a change of variables in the integral appearing in 
Eq.~(\ref{Eq:Coupled_SF_PP2}). 
Once the values of $\varv$ and $\theta_\varv$ (the velocity vector appearing 
as the argument of the density matrix element for which we write the 
statistical equilibrium equation) and the direction $\vec \Omega$ (through 
the angles $\theta$ and $\chi$) are specified, the velocity component 
$\varv_{\rm c}$ is fixed. 
We can then transform the double integral in the variables 
$({\rm d} \varv'_{\rm a}, {\rm d} \varv'_{\rm b})$ in a double integral over 
the variables $(\varv'_{\ast},\theta_{\ast})$. 
For this we have to consider the formal transformation
\begin{equation}
   \int \frac{{\rm d} \varv_{\rm a}^\prime}{\varv_{\rm t}} 
   \int \frac{{\rm d} \varv_{\rm b}^\prime}{\varv_{\rm t}} 
   \rightarrow \frac{1}{\varv_{\rm t}^2} \, \int {\rm d} \varv'_{\ast}
   \int {\rm d} (\cos \theta_*) \, \frac{1}{|\, {\mathcal J}\, |} \; , 
   \label{Eq:int_transf}
\end{equation}
where ${\mathcal J}$ is the determinant of the Jacobian of the transformation, 
namely
\begin{equation}
   {\mathcal J} = {\rm det} 
   \left( \begin{array}{cc} 
    \displaystyle{\frac{{\rm d} \varv'_{\ast}}{{\rm d} \varv'_{\rm a}}} &
    \displaystyle{\frac{{\rm d} \varv'_{\ast}}{{\rm d} \varv'_{\rm b}}} \\
     & \\
    \displaystyle{\frac{{\rm d} (\cos\theta_{\ast})}{{\rm d} \varv'_{\rm a}}} &
    \displaystyle{\frac{{\rm d} (\cos\theta_{\ast})}{{\rm d} \varv'_{\rm b}}}
	  \end{array}
   \right) \; .
\end{equation}
On the other hand, from the equations relating $\varv'_{\ast}$ and 
$\theta_{\ast}$ with $\varv'_{\rm a}$ and $\varv'_{\rm b}$, we have 
\begin{equation}
   \frac{{\rm d} \varv'_{\ast}}{{\rm d} \varv'_{\rm a}} = 
   \frac{\varv'_{\rm a}}{\varv'_{\ast}} \; , \qquad
   \frac{{\rm d} \varv'_{\ast}}{{\rm d} \varv'_{\rm b}} = 
   \frac{\varv'_{\rm b}}{\varv'_{\ast}} \; , \qquad
   \frac{{\rm d} (\cos \theta_{\ast})}{{\rm d} \varv'_{\rm a}} = 
   \frac{-\sin \theta}{\varv'_{\ast}} \; , \qquad 
   \frac{{\rm d} (\cos \theta_{\ast})}{{\rm d} \varv'_{\rm b}} = 0 \; ,
\end{equation}
so that we obtain
\begin{equation}
   \frac{1}{|\, {\mathcal J}\, |} = \frac{\varv_{\ast}^{\prime 2}}{\sin \theta 
   \, | \, \varv'_{\rm b} \, |} \; .
\end{equation}
This equation, together with the relationships previously developed among the 
components of the velocity, allow to rewrite Eq.~(\ref{Eq:int_transf}) in the 
form
\begin{equation}
   \int \frac{{\rm d} \varv_{\rm a}^\prime}{\varv_{\rm t}} 
   \int \frac{{\rm d} \varv_{\rm b}^\prime}{\varv_{\rm t}} 
   \rightarrow  
   \int \frac{{\rm d} \varv'_{\ast}}{\varv_{\rm t}} \,
   \frac{\varv'_{\ast}}{\varv_{\rm t}} \int {\rm d} \theta_{\ast} 
   \frac{1}{\sin \theta \, |\sin (\chi_{\ast} - \chi) \,|} \; ,
\end{equation}
where
\begin{equation}
   |\sin (\chi_{\ast} - \chi) \,| = \sqrt{1 - \left( \frac{\varv_{\rm c} - 
   \varv'_{\ast} \, \cos \theta_{\ast} \, \cos \theta}{\varv'_{\ast} \, 
   \sin \theta_{\ast} \, \sin \theta } \right)^2 } \; .
\end{equation}
Substituting this result into Eq.~(\ref{Eq:Coupled_SF_PP2}), and inverting the 
order of the integrals, it is possible to rewrite the same equation in a 
different form. 
Performing the following change of notations on the integration variables: 
$\varv'_{\ast} \rightarrow \varv'$, 
$\theta_{\ast} \rightarrow \theta'_\varv$, one gets
\begin{equation}
	\left[ 1+ \epsilon+ \delta_{\,u}^{\, (K)}(\boldsymbol{\varv} \,) \right]
   	{\mathcal S}^K_Q(\varv, \theta_\varv, 0, t_{\rm L}) = 
   	\delta_{K0} \, \delta_{Q0} \; \epsilon \, B_T(\nu_0) 
   	+ \int_0^{\infty} \! {\rm d} t'_{\rm L} \; 
   	\int \frac{{\rm d} \varv'}{\varv_{\rm t}} \, 
	\frac{\varv'}{\varv_{\rm t}} \int {\rm d} \theta'_\varv \sum_{K' Q'}
   	{\mathcal A}^{K K'}_{Q Q'} (\varv, \theta_\varv, t_{\rm L}, \varv', 
   	\theta'_\varv, t'_{\rm L}) \, {\mathcal S}^{K'}_{Q'} \! (\varv', 
   	\theta'_\varv, 0, t'_{\rm L}) \; ,
\end{equation}
where the kernel, ${\mathcal A}^{K K'}_{Q Q'} (\varv, \theta_\varv, t_{\rm L}, 
\varv', \theta'_\varv,t'_{\rm L})$ is given by
\begin{equation}
   {\mathcal A}^{K K'}_{Q Q'} (\varv, \theta_\varv, t_{\rm L}, \varv', 
   \theta'_\varv, t'_{\rm L}) = \frac{1}{4 \pi} \, \frac{1}{\pi^{3/2}}
   \int_0^{2 \pi} \! {\rm d}\chi \int_{\theta_1}^{\theta_2} \! 
   {\rm d} \theta \, \frac{1}{|\cos \theta \,|} \;
   {\rm e}^{- (t'_{\rm L} - t_{\rm L}) \, \varphi(\xi') / \cos \theta} \,
   {\rm e}^{-(\varv' / \varv_{\rm t})^2} \, w^{\, (K)}_{J_u J_\ell} \;
   w^{\, (K')}_{J_u J_\ell} \, \Gamma_{K Q, K' Q'} (\vec \Omega) \, 
   \frac{{\rm e}^{-{\rm i}\, Q'\chi_{\ast}}}{|\sin (\chi_{\ast} - \chi) \,|}
\end{equation}
and can be more conveniently expressed in the form
\begin{equation}
   {\mathcal A}^{K K'}_{Q Q'} (\varv, \theta_\varv, t_{\rm L}, \varv', 
   \theta'_\varv, t'_{\rm L}) = \frac{1}{4 \pi} \, \frac{1}{\pi^{3/2}} 
   \int_0^{2 \pi} \! {\rm d}\chi \int_{\theta_1}^{\theta_2} \! {\rm d}\theta \, 
   \frac{1}{|\cos \theta \,|} \;
   {\rm e}^{-(t'_{\rm L} - t_{\rm L}) \, \varphi(\xi') / \cos \theta} \;
   {\rm e}^{-(\varv' / \varv_{\rm t})^2}  w^{\, (K)}_{J_u J_\ell} \; 
   w^{\, (K')}_{J_u J_\ell} \, \Gamma_{K Q, K' Q'} (\vec \Omega) \, 
   {\rm e}^{-{\rm i} \, Q' \chi} \, 
   \frac{{\rm e}^{-{\rm i} \, Q'(\chi_{\ast}-\chi)}}
   {|\sin (\chi_{\ast}-\chi) \,|} \; .
   \label{Eq:kernel_PP}
\end{equation}
The quantities $\cos(\chi_{\ast} - \chi)$ and $\sin(\chi_{\ast} - \chi)$ 
appearing (implicitly or explicitly) in this equation are related to the 
different variables through the equations
\begin{align}
   \cos(\chi_{\ast} -\chi) & = \frac{\varv_{\rm c} - \varv' \, 
   \cos \theta'_\varv \, \cos \theta}{\varv' \, \sin \theta'_\varv \, \sin 
   \theta} \; , \\
   \sin(\chi_{\ast} -\chi) & = \pm \sqrt{1 - \left( 
   \frac{\varv_{\rm c} - \varv' \, \cos \theta'_\varv \, \cos \theta}
   {\varv' \, \sin \theta'_\varv \, \sin \theta} \right)^2 } \; .
\end{align}
Concerning the $\pm$ sign appearing in this last expression, it is important to 
note that it is connected with the sign of the velocity component,
$\varv'_{\rm b}$, that has now disappeared from the equations, due to change of
variables that we have performed. 
Equation~(\ref{Eq:kernel_PP}) has then, more properly, to be written as
\begin{equation}
\begin{split}
   {\mathcal A}^{K K'}_{Q Q'}(\varv, \theta_\varv, t_{\rm L}, \varv', 
   \theta'_\varv, t'_{\rm L}) = \frac{1}{4 \pi} \, \frac{1}{\pi^{3/2}} 
   \int_0^{2 \pi} \! {\rm d}\chi \int_{\theta_1}^{\theta_2} \! 
   {\rm d}\theta & \, \frac{1}{|\cos \theta \, |} \;
   {\rm e}^{-(t'_{\rm L} - t_{\rm L}) \, \varphi(\xi') / \cos \theta} \;
   {\rm e}^{-(\varv' / \varv_{\rm t})^2} w^{\, (K)}_{J_u J_\ell} \; 
   w^{\, (K')}_{J_u J_\ell} \, \times \\
   & \times \Gamma_{K Q, K' Q'} (\vec \Omega) \, {\rm e}^{-{\rm i} \, Q' \chi} 
   \, \frac{1}{|\sin (\chi_{\ast} - \chi) \,|}
   \left[  {\rm e}^{-{\rm i}\, Q' (\chi_{\ast}^{(-)}-\chi)} 
   + {\rm e}^{-{\rm i}\, Q' (\chi_{\ast}^{(+)}-\chi)} \right] \; , 
\end{split}
\end{equation}
where  $\chi_{\ast}^{(-)}$ and $\chi_{\ast}^{(+)}$ are the values of 
$\chi_{\ast}$ corresponding, respectively, to negative or positive 
$\varv'_{\rm b}$. With easy algebra we find
\begin{align}
   \left[ {\rm e}^{-{\rm i} \, Q' (\chi_{\ast}^{(-)}-\chi)} 
   + {\rm e}^{-{\rm i} \, Q' (\chi_{\ast}^{(+)}-\chi)} \right] & =
   2 \;\; ,
   \qquad \qquad \qquad \qquad \qquad {\rm for \;} Q'=0 \;\; , \nonumber \\
   \left[ {\rm e}^{-{\rm i} \, Q' (\chi_{\ast}^{(-)}-\chi)} 
   + {\rm e}^{-{\rm i} \, Q' (\chi_{\ast}^{(+)}-\chi)} \right] & =
   2 \cos(\chi_{\ast} -\chi) \;\; ,
   \qquad \qquad \quad \, {\rm for \;} Q' = \pm 1 \;\; , \\
   \left[ {\rm e}^{-{\rm i} \, Q' (\chi_{\ast}^{(-)}-\chi)} 
   + {\rm e}^{-{\rm i} \, Q' (\chi_{\ast}^{(+)}-\chi)} \right] & =
   2 \, [ 2 \, \cos^2(\chi_{\ast} -\chi) - 1] \;\; ,
   \qquad \! {\rm for \;} Q' = \pm 2 \nonumber \;\; .
\end{align}

We conclude observing that the equations for the velocity-independent source 
function derived in Chapter~14 of LL04 under the approximation of complete 
redistribution on velocity can be recovered, as a particular case, from the 
equations presented in this work.
Indeed, they can be obtained starting from Eq.~(\ref{Eq:Coupled_SF2}) 
(or one of its following reformulations for the case of a plane-parallel 
atmosphere), assuming that the source function ${\mathcal S}^K_Q$ and the 
depolarizing rate $\delta_u^{(K)}$ do not depend on the velocity, integrating 
over the velocity components $\varv_a^{\prime}$ and $\varv_b^{\prime}$, and 
averaging over the Maxwellian distribution of the velocity component $\varv_c$.

\section{The $R_{\rm I}$ redistribution phase-matrix}
We now analyze the basic equations that underly this physical problem, 
following an alternative approach.
Instead of eliminating the ``radiation field variables'' in order to obtain a 
set of coupled equations for the velocity-dependent density matrix elements, 
we can eliminate the ``density matrix variables'' in order to obtain equations 
which directly involve the Stokes parameters of the radiation field. 
This will bring us to the definition of a suitable redistribution phase-matrix, 
referred to in the literature as $R_{\rm I}$ in the case of a Maxwellian 
distribution of velocities.

Neglecting the magnetic field contribution, Eq.~(\ref{Eq:SF_SEE}) can be 
easily solved for ${\mathcal S}^K_Q(\boldsymbol{\varv}, \vec x)$. 
One gets
\begin{equation}
	{\mathcal S}^K_Q(\boldsymbol{\varv}, \vec x\,) = 
	\frac{w^{\, (K)}_{J_u J_\ell} \; (-1)^Q \, 
	[ J^K_{-Q}(\nu_0)]_{\rm c.f.} + \delta_{K0} \, \delta_{Q0} \; 
   	\epsilon \, B_T(\nu_0)}{1+ \epsilon + 
	\delta_{\,u}^{\, (K)}(\boldsymbol{\varv} \,)} \; .
\end{equation}
We can now substitute this expression for the velocity-dependent source 
function into Eq.~(\ref{Eq:emis_coef}), giving the emission coefficient for 
the Stokes parameters at frequency $\nu$ into the direction $\vec \Omega$. 
Taking also into account Eqs.~(\ref{Eq:JKQ_cf}) and (\ref{Eq:abs_prof}), the 
expression of the emission coefficient can be cast in the form
\begin{equation}
	\varepsilon_i(\nu, \vec \Omega) = k_{\rm L}^{\rm A}(\vec{x}) 
    	\left\{ \frac{\epsilon}{1 + \epsilon} \, p(\nu,\vec{\Omega}) \, 
	B_T(\nu_0) + \frac{1}{1 + \epsilon} \int {\rm d}^3 
	\boldsymbol{\varv} \, f(\boldsymbol{\varv} \,) \, \delta 
	\left( \nu_0 + \nu_0 \frac{\boldsymbol{\varv} \cdot \vec \Omega}{c} 
	- \nu \right) \oint \frac{{\rm d} \vec \Omega'}{4 \pi}
    	\sum_{j=0}^3 P_{ij}^{({\rm c})}(\vec \Omega, \vec \Omega', 
	\boldsymbol{\varv}) \, S_{\! j} \left( \nu_0 + \nu_0 
	\frac{\boldsymbol{\varv} \cdot \vec \Omega'}{c}, \vec\Omega' \right) 
	\right\} \; , 
\label{Eq:eps_red1}
\end{equation}
where $P_{ij}^{({\rm c})}(\vec \Omega, \vec \Omega', \boldsymbol{\varv})$ is 
the scattering phase matrix (corrected for depolarizing collisions) which is 
given by
\begin{equation}
	P_{ij}^{({\rm c})}(\vec \Omega, \vec \Omega', \boldsymbol{\varv}) = 
	\sum_{KQ} \left[ 1+ \frac{\delta_u^{(K)}(\boldsymbol{\varv} \, )}
	{1 + \epsilon} \right]^{-1} W_K(J_\ell,J_u) \, (-1)^Q \, 
	{\mathcal T}^K_Q (i, \vec \Omega) \, 
	{\mathcal T}^K_{-Q}(j,\vec \Omega') \; ,
\end{equation}
with $W_K(J_{\ell},J_u) = \left( w^{(K)}_{J_u J_{\ell}} \right)^2$. 
We now assume that the velocity distribution is Maxwellian, and we perform the 
integral in ${\rm d}^3 \boldsymbol{\varv}$, neglecting the velocity dependence
of the depolarizing collisions\footnote{This is a non-trivial approximation. 
Indeed, if the colliding hydrogen atoms have an average quadratic velocity $w$, 
an atom that is moving with velocity $\boldsymbol{\varv}$ ``sees'', in his 
rest frame, that the hydrogen atoms have an average quadratic velocity 
$\sqrt{\varv^2 + w^2}$. 
This can lead to an important dependence of the quantities $\delta^{(K)}$ on 
$\boldsymbol{\varv}$.}. 
Under this hypothesis, the scattering phase matrix does not depend any 
longer on $\boldsymbol{\varv}$ and can thus be simply written as 
$P_{ij}^{({\rm c})}(\vec \Omega, \vec \Omega')$.
It is convenient to perform an inversion of the two integrals (first we 
perform the integral in ${\rm d^3} \boldsymbol{\varv}$ and then the one in 
${\rm d} \Omega'$) and to introduce a right-handed triplet of unit vectors, 
$(\vec u_1, \vec u_2, \vec u_3)$, defined in the following way: $\vec u_1$ is 
the unit vector along the direction that bisects the angle 
$\Theta$  $(0 \le \Theta \le \pi)$ formed by $\vec \Omega$ and 
$\vec \Omega'$; $\vec u_2$ is perpendicular to $\vec u_1$, lying, as 
$\vec u_1$, in the plane defined by $\vec \Omega$ and $\vec \Omega'$, 
and being directed in such a way that it has a positive component along 
$\vec \Omega$ . Finally, $\vec u_3$ is defined accordingly.  
The unit vectors $\vec u_1$ and $\vec u_2$ are given by
\begin{equation}
   \vec u_1 = \frac{\vec \Omega + \vec \Omega'}{2 \, \cos(\Theta/2)} \;\; , 
   \qquad \vec u_2 = \frac{\vec \Omega - \vec \Omega'}{2 \, \sin(\Theta/2)} 
   \; ,
\end{equation}
with the inverse formulae
\begin{equation}
   \vec \Omega = \cos(\Theta/2) \, \vec u_1 + \sin(\Theta/2) \, \vec u_2 \;\; ,
   \qquad \vec \Omega' = \cos(\Theta/2) \, \vec u_1 - \sin(\Theta/2) \, 
   \vec u_2 \; .
\end{equation}
Writing $ \boldsymbol{\varv} = \varv_1 \, \vec u_1 + \varv_2 \, \vec u_2 + 
\varv_3 \, \vec u_3$, the delta function appearing in Eq.~(\ref{Eq:eps_red1}) 
is satisfied when
\begin{equation}
   \frac{c}{\nu_0} \, (\nu - \nu_0) = \cos(\Theta/2) \, \varv_1 + 
   \sin(\Theta/2) \, \varv_2  \; .
   \label{Eq:int_delta}
\end{equation}
Taking into account this relation between $\varv_1$ and $\varv_2$, the Stokes 
parameter $S_{\! j}(\nu_0 + \nu_0 \, \boldsymbol{\varv} \cdot \vec\Omega' /c, 
\vec\Omega')$, also appearing in Eq.~(\ref{Eq:eps_red1}), can be written as 
$S_{\! j}(\nu',\vec \Omega')$, where the frequency $\nu'$ only depends on 
$\varv_2$, being given by
\begin{equation}
	\nu' = \nu -2 \frac{\nu_0}{c} \, \sin(\Theta/2) \, \varv_2  \;\ .
\label{Eq:nuprime}
\end{equation}
Integrating first in ${\rm d} \varv_3$ and then in ${\rm d} \varv_1$ (this 
latter integral being performed taking into account the Dirac-delta), one
is left with the expression
\begin{equation}
   \varepsilon_i(\nu, \vec \Omega) = k_{\rm L}^{\rm A}(\vec{x})
   \left\{ \frac{\epsilon}{1 + \epsilon} \, p_{\rm M}(\nu - \nu_0) \, 
   B_T(\nu_0) + \frac{1}{1 + \epsilon} \oint \frac{{\rm d} \vec \Omega'}{4 \pi} 
   \sum_{j=0}^3 P_{ij}^{({\rm c})} (\vec \Omega, \vec \Omega') 
   \frac{1}{\pi \, \Delta \nu_{\rm D} \, \cos(\Theta/2)} \, 
   \int \frac{{\rm d} \varv_2}{\varv_{\rm t}} \, 
   {\rm e}^{-(\varv_1^2 + \varv_2^2)/\varv_{\rm t}^2} \, 
   S_{\! j} (\nu', \vec \Omega') \right\} \; , 
   \label{Eq:eps_red2}
\end{equation}
where $\varv_1$ follows from Eq.~(\ref{Eq:int_delta}), being given by
\begin{equation}
   \varv_1= \frac{1}{\cos(\Theta/2)} \left( \frac{c}{\nu_0} \, 
   (\nu - \nu_0) - \sin(\Theta/2) \, \varv_2 \right) \; .
   \label{Eq:v1}
\end{equation}

This equation can be written in an alternative form by a change of variable 
in the second integral, passing from the variable $\varv_2$ to the variable 
$\nu'$. 
This can be done by taking into account that (see Eq.~(\ref{Eq:nuprime}))
\begin{equation}
   \frac{{\rm d} \varv_2}{\varv_{\rm t}} = -\frac{1}{2 \, \Delta \nu_{\rm D} \,    \sin(\Theta/2)} \, {\rm d}\nu' \; ,
\end{equation}
and that, starting from Eqs.~(\ref{Eq:nuprime}) and (\ref{Eq:v1}), the quantity 
$(\varv_1^2 + \varv_2^2)/\varv_{\rm t}^2$ can be written, after some algebra, 
in the form
\begin{equation}
   \frac{\varv_1^2 + \varv_2^2}{\varv_{\rm t}^2} = 
   \frac{1}{\Delta \nu_{\rm D}^2 \sin^2 \! \Theta} 
   \left[ (\nu - \nu_0)^2 + (\nu' - \nu_0)^2 - 2 \, (\nu-\nu_0)
   (\nu' - \nu_0) \, \cos \! \Theta \right] \; .
\end{equation}
Moreover, by introducing the reduced variables $\xi$ and $\xi'$, defined by
\begin{equation}
   \xi = \frac{\nu - \nu_0}{\Delta \nu_{\rm D}} \;\; , \qquad
   \xi'= \frac{\nu' - \nu_0}{\Delta \nu_{\rm D}} \; ,
\end{equation}
taking into account that $\varepsilon_i(\xi,\vec{\Omega}) = 
\varepsilon_i(\nu,\vec{\Omega}) \, {\rm d} \nu/{\rm d} \xi = 
\varepsilon_i(\nu,\vec{\Omega}) \, \Delta \nu_{\rm D}$, and recalling 
the definition of the absorption profile $\varphi(\xi)$ (see 
Eq.~(\ref{Eq:abs_prof_red})), Eq.~(\ref{Eq:eps_red2}) can be written in the 
following form
\begin{equation}
   \varepsilon_i(\xi, \vec \Omega) = k_{\rm L}^{\rm A}(\vec{x})
   \left\{ \frac{\epsilon}{1 + \epsilon} \, \varphi(\xi) \, B_T(\nu_0) + 
   \frac{1}{1 + \epsilon} \oint \frac{{\rm d} \vec \Omega'}{4 \pi} 
   \sum_{j=0}^3 P_{ij}^{({\rm c})} (\vec \Omega, \vec \Omega') 
   \frac{1}{\pi \, \sin \! \Theta} \, \int {\rm d}\xi' \, 
   {\rm e}^{-(\xi^2 + \xi^{\prime 2} - 2 \, \xi \xi' \cos \! \Theta)
   / \! \sin^2 \Theta} \, S_{\! j} (\xi', \vec \Omega') \right\}  
\end{equation}
or, using the lexicon of `redistribution functions',
\begin{equation}
   \varepsilon_i(\xi, \vec \Omega) = k_{\rm L}^{\rm A}(\vec{x})
   \left\{ \frac{\epsilon}{1 + \epsilon} \, \varphi(\xi) \, B_T(\nu_0) + 
   \frac{1}{1 + \epsilon} \oint \frac{{\rm d} \vec{\Omega'}}{4 \pi}  
   \int {\rm d}\xi' \, \sum_{j=0}^3 
   \left[ R_{\rm I}(\xi,\vec \Omega,\xi',\vec \Omega') \right]_{ij} \, 
   S_{\! j} (\xi', \vec \Omega') \right\} \; , 
\end{equation}
where
\begin{equation}
   \left[ R_{\rm I}(\xi,\vec \Omega,\xi',\vec \Omega') \right]_{ij}  
   = \frac{1}{\pi \sin \Theta} P_{ij}^{({\rm c})} (\vec \Omega, \vec \Omega') 
   \, {\rm e}^{-(\xi^2 + \xi^{\prime 2} - 2 \, \xi \xi' \cos \! \Theta)
   / \! \sin^2 \! \Theta} \; .
   \label{Eq:RI}
\end{equation}
This redistribution matrix was first proposed heuristically by 
\citet{Dum77}.
We also observe that for a $0 \rightarrow 1$ transition (for which 
$W_0=W_2=1$), and in the absence of depolarizing collisions 
($\delta_u^{(K)}=0$), the redistribution matrix element $[R_I]_{00}$ 
corresponds to the redistribution function $R_I$ derived by \citet{Hum62} for 
the unpolarized case (see his Eqs.~(2.21.2) and 
(2.21.4)).\footnote{The discrepancy by a factor $16 \pi^2$ is due to the 
different normalization of the redistribution functions.}

\section{Conclusions}
In this paper we have derived the equations for the Non-LTE problem of the 
$2^{\rm nd}$ kind, taking velocity density matrix correlations into account. 
We considered the basic case of a two-level atom with infinitely sharp upper 
and lower levels, and we derived the statistical equilibrium equations for the 
velocity-space density matrix, neglecting the generalized Boltzmann term (which 
is a good approximation in the outer layers of a stellar atmosphere).
Taking the Doppler effect into account, we derived a set of coupled equations 
for the velocity-dependent multipole components of the source function.
Such equations show a coupling between the atoms at point $\vec x$, moving with 
velocity $\boldsymbol{\varv}$, and the atoms at point $\vec x^{\prime}$, moving 
with velocity $\boldsymbol{\varv}^{\prime}$, such that the difference 
$\boldsymbol{\varv} - \boldsymbol{\varv}^{\prime}$ (the relative velocity) is 
perpendicular to the direction $\vec x - \vec x^{\prime}$.
This is a clear consequence of the Doppler effect and of the fact that we 
considered a resonant transition between two infinitely sharp atomic levels, 
an academic case that nevertheless allows to investigate in detail the physics 
of pure Doppler redistribution in scattering polarization.

Starting from the statistical equilibrium equations for the velocity-space 
density matrix, and from the radiative transfer equations for polarized 
radiation, we derived the redistribution matrix corresponding to this physical 
problem.
This redistribution matrix \citep[heuristically proposed by][]{Dum77} provides 
an alternative, equivalent description of the problem, and represents the 
generalization to the polarized case of the (angle-dependent) $R_I$ 
redistribution function derived by \citet{Hum62}.
If, on the one hand, the redistribution matrix formalism allows a very simple 
and intuitive description of redistribution phenomena, on the other hand, the 
velocity-space density matrix formalism provides a very transparent picture
of the physics of the atom-photon interaction.
The above-mentioned correlations between atoms located at different points of 
the plasma remain actually ``hidden'' in the redistribution matrix formalism.
It should also be observed that only the average effect of such correlations is 
taken into account if the approximate ``angle-averaged'' $R_{I}^{AA}$ 
redistribution matrix is considered.
The velocity-space density matrix formalism has also the advantage of being 
suitable for taking lower-level polarization into account, and for describing 
multilevel atomic systems.
An important point to remark is the appearance, in the statistical equilibrium 
equations for the velocity-space density matrix, of the generalized Boltzmann 
term. 
This term allows to include the effect of velocity-changing collisions into the 
problem, and points out the limits of applicability of any theoretical approach 
in which it is neglected.
Velocity-changing collisions are a complex and not yet deeply investigated 
physical aspect that, however, may play an important role in the lower layers 
of stellar atmospheres.
We point out that the equations that have been derived in Sect.~5 (in 
particular Eq.~(\ref{Eq:Coupled_SF2})) are very general, and can be applied 
to arbitrary velocity distributions and to plasma structures of any geometry.

\begin{acknowledgements}
Financial support by the Spanish Ministry of Economy and Competitiveness
through project AYA2010-18029 (Solar Magnetism and Astrophysical
Spectropolarimetry) is gratefully acknowledged.
One of the authors (ELD) wishes to acknowledge the Research Area of the
Instituto de Astrof\'isica de Canarias (IAC) for helping to finance a six
months sabbatical leave at the IAC, during which part of this work was carried
out.
\end{acknowledgements}

\end{document}